\documentclass[a4paper,fleqn,usenatbib]{mnras}

\usepackage{newtxtext,newtxmath}

\usepackage[T1]{fontenc}
\usepackage{ae,aecompl}
\usepackage{subfig}
\usepackage{comment}
\usepackage{xspace}
\usepackage{tikz,pgf}

\newcommand{\degrees}{\mbox{$^{\circ}$}\xspace}
\newcommand{\vsini}{\mbox{$\varv\sin i$}\xspace}
\newcommand{\Prot}{\mbox{$P_{\mathrm{rot}}$}\xspace}
\newcommand{\Rstar}{\mbox{$R_{\ast}$}\xspace}
\newcommand{\numax}{\mbox{$\nu_{\mathrm{max}}$}\xspace}

\usepackage{graphicx}	
\usepackage{amsmath}	
\usepackage{amssymb}	

\usepackage{booktabs}
\usepackage{tabularx}
\usepackage{float}

\title[Stellar Inclination Angles]{Bayesian hierarchical inference of asteroseismic inclination angles}

\author[J. S. Kuszlewicz et al.]
{James S. Kuszlewicz,$^{1,2,3}$\thanks{E-mail: kuszlewicz@mps.mpg.de}
William J. Chaplin,$^{2,3}$
Thomas S. H. North,$^{2,3}$
\newauthor
Will M. Farr,$^{4,5,6}$
Keaton J. Bell,$^{1,3}$
Guy R. Davies,$^{2,3}$
Tiago L. Campante,$^{7,8}$
\newauthor
and Saskia Hekker$^{1,3}$
\\
$^{1}$ Max-Planck-Institut f\"{u}r Sonnensystemforschung, Justus-von-Liebig-Weg 3, 37077 G\"{o}ttingen, Germany \\
$^{2}$School of Physics and Astronomy, University of Birmingham, Edgbaston, Birmingham, B15 2TT, UK\\
$^{3}$Stellar Astrophysics Centre (SAC), Department of Physics and Astronomy, Aarhus University, Ny Munkegade 120, \\ DK-8000 Aarhus C, Denmark\\
$^{4}$Center for Computational Astrophysics, Flatiron Institute, 162 Fifth Avenue, New York, NY 10010, USA\\
$^{5}$Department of Physics and Astronomy, Stony Brook University, Stony Brook, NY 11794-3800, USA\\
$^{6}$School of Physics and Astronomy and Birmingham Institute of Gravitational Wave Astronomy, University of Birmingham, \\ Birmingham B15 2TT, UK\\
$^{7}$Instituto de Astrof\'{i}sica e Ci\^{e}ncias do Espa\c{c}o, Universidade do Porto, Rua das Estrelas, P-4150-762 Porto, Portugal\\
$^{8}$Departamento de F\'{i}sica e Astronomia, Faculdade de Ci\^{e}ncias da Universidade do Porto, Rua do Campo Alegre, s/n, \\ P-4169-007 Porto, Portugal
}

\date{Accepted XXX. Received YYY; in original form ZZZ}

\pubyear{2018}

\begin{document}
\label{firstpage}
\pagerange{\pageref{firstpage}--\pageref{lastpage}}
\maketitle
\begin{abstract}
The stellar inclination angle---the angle between the rotation axis of a star and our line of sight---provides valuable information in many different areas, from the characterisation of the geometry of exoplanetary and eclipsing binary systems, to the formation and evolution of those systems. We propose a method based on asteroseismology and a Bayesian hierarchical scheme for extracting the inclination angle of a single star. This hierarchical method therefore provides a means to both accurately and robustly extract inclination angles from red giant stars. We successfully apply this technique to an artificial dataset with an underlying isotropic inclination angle distribution to verify the method. We also apply this technique to 123 red giant stars observed with \emph{Kepler}. We also show the need for a selection function to account for possible population-level biases, that are not present in individual star-by-star cases, in order to extend the hierarchical method towards inferring underlying population inclination angle distributions.

\end{abstract}

\section{Introduction}

Traditionally, the determination of the inclination angle, $i$, of a star requires knowledge of the rotation period of the star, \Prot, along with the stellar radius, \Rstar, and projected equatorial rotational velocity, \vsini (e.g. \citealt{1994AJ....107..306H, 2007AJ....133.1828W,2010ApJ...719..602S, 2012ApJ...756...66H}) 
\begin{equation}
    i = \arcsin\left[\frac{\varv \sin i}{\left(2\pi R_{\ast}/P_{\mathrm{rot}}\right)}\right].
\label{eqn:trad_i}
\end{equation}
This technique has been applied to both main-sequence and red-giant stars where such measurements have been available and can offer constraints on the inclination angle \citep{2010ApJ...719..602S, 2012ApJ...756...66H,2015ApJ...807...82T,2017A&A...605A.111C}. Whilst this is not generally a problem for fast rotators for which \Prot or \vsini measurements can be attained, obtaining a measurement of the rotation period or \vsini for slower rotators can prove to be complicated. A difficulty with this method for calculating the inclination angle is that it can lead to unphysical solutions where $\sin i > 1$, which implies that one (or possibly more) of the parameters have been incorrectly inferred, in the context of Eqn~\ref{eqn:trad_i}. The probabilistic method introduced by \cite{2014ApJ...796...47M} aimed to address this by deriving the $\cos i$ distribution given the \vsini, \Prot and radius \Rstar distributions. This subsequent technique has been successfully applied to a number of systems and used in larger ensemble analyses \citep{2014ApJ...796...47M, Tiago}.

Asteroseismology offers a means to measure the inclination angle, through the non-radial oscillation modes (e.g. \citealt{1985ApJ...292..238P, 2003ApJ...589.1009G, 2013ApJ...766..101C, 2018arXiv180507044K}). The advantage of asteroseismology is that it can be applied to stars that are relatively slow rotators for which \vsini or \Prot measurements are difficult to obtain. This provides a great opportunity to measure the inclination angle that does not explicitly require the acquisition of \vsini, \Prot or \Rstar. Asteroseismology has been used to extract the inclination angle in a number of different situations, such as transiting exoplanetary systems \citep{2013ApJ...766..101C, 2013Sci...342..331H,2014A&A...570A..54L} and in clusters (e.g. \citealt{2017NatAs...1E..64C}). This is the approach we adopt in this work.

There have been many studies where the inclination angle has been an important component, such as in the characterisation of exoplanet and eclipsing binary systems, as it can help reveal the underlying geometry of the system through constraints on the obliquity, the angle between the orbital axis and the stellar rotation axis (e.g. \citealt{1995AcA....45..725G, 2010ApJ...719..602S, 2012ApJ...756...66H,2014ApJ...783....9H,2014ApJ...796...47M,2015MNRAS.446.2959D,Tiago}). In the case of Kepler-56 \citep{2013Sci...342..331H}, the asteroseismically determined inclination angle showed that the system is misaligned (the plane of the planetary orbits are not perpendicular to the rotation axis of the star). As of now, this is the only known misaligned multi-planet system with an evolved host.

In addition to looking at the inclination angle of individual systems, the analysis of an ensemble of inclination angle (or $\sin i$, where $i$ is the inclination angle) measurements is also extremely valuable. By looking at the distribution at a population level, i.e. the distribution of a large sample of stars \citep{2001AJ....122.2008A}, it is possible to test fundamental assumptions used in astrophysical analyses. One such assumption is that the distribution of stellar inclination angles is random (isotropic) with respect to our line of sight \citep{1950ApJ...111..142C}. This is subsequently used in analyses when simulated data are required. For example, the calculation of simulated \vsini values in \cite{2010ApJ...719..602S} involved this assumption to construct the null hypothesis used to test spin-orbit alignment in exoplanet systems. It is also possible to shed light on cluster formation processes through the ensemble analysis of inclination angles. \cite{2010MNRAS.402.1380J} derived the distribution of $\sin i$ values for stars in the open clusters Pleiades and Alpha Per from \Prot, \vsini and radius values derived from the cluster distances, and found no evidence of spin-orbit alignment in either cluster using a Monte-Carlo modelling technique. 

The difficulty of extracting the inclination angle has been highlighted by the disagreement in the asteroseismic analyses by \cite{2017NatAs...1E..64C} and \cite{2018arXiv180708301M} when analysing the \emph{Kepler} \citep{2010Sci...327..977B} clusters NGC 6819 and NGC 6791 \citep{2010ApJ...713L.182S, 2011A&A...530A.100H, 2011ApJ...729L..10B, 2011ApJ...739...13S}. \cite{2017NatAs...1E..64C} reported that both clusters showed evidence of strong alignment in their inclination angle distributions, suggesting that the global angular momentum of the initial gas clouds during the cluster formation process were efficiently transferred to the stars leading to a detectable imprint in the inclination angles. However, \cite{2018arXiv180708301M} also analysed these clusters and found that there was no such evidence for strong alignment and in fact that both were consistent with being randomly distributed. This difference in the derived inclination angle arises due to the possibility of misinterpreting whether a star is an extremely slow rotator or possesses a low inclination angle (see section \ref{sec:inc}), which is currently unresolved in these cases.

In this work we follow the method put forward by \cite{2010ApJ...725.2166H} and adopt a Bayesian approach to infer the underlying inclination angle of a set of stars from asteroseismic estimates. 

\section[]{Data}

Long-cadence data from the \emph{Kepler} mission are used (a cadence of 29.4 minutes) with all 4 years of observations from quarters Q1-Q17. Our sample consists of 123 stars taken from the 13,000 red giants in \cite{2013ApJ...765L..41S} with a $\nu_{\mathrm{max}}$ (determined by \citealt{2013ApJ...765L..41S}) in the range 231-274 $\mu$Hz. This \numax range was chosen because the probability of a star possessing overlapping modes (due to rapid rotation) is very low, therefore making the identification of the mixed modes (those with a mixed p- and g-character, see e.g. \citealt{2014A&A...572L...5M}) easier. All photometric timeseries were produced using the pipeline developed by \cite{2010ApJ...713L.120J}, and power spectra were obtained using the Lomb-Scargle periodogram (\citealt{1976Ap&SS..39..447L,1982ApJ...263..835S}). 

Due to the complexity of red giant oscillation spectra, not all of the 123 red giants in our sample could be used in the analysis. A few were observed to show suppressed $\ell=1$ modes \citep{2014A&A...563A..84G, 2015Sci...350..423F, 2016Natur.529..364S, 2017A&A...598A..62M}, which greatly hinder our ability to extract the inclination angle. There were also some cases where the stars had oscillation modes reflected from across the Nyquist frequency \citep{2014MNRAS.445..946C}. This made the mode identification difficult because there was a mixture of real and reflected oscillation modes obscuring the well established frequency patterns. Finally, there were also a few stars for which the modes could not be successfully disentangled with the current method due to the highly complex spectra. 

\section[]{Data Analysis}
\label{sec:analysis}

\subsection{Mode Detection}\label{sec:det}

The first step towards extracting the inclination angle is to detect the $\ell=1$ mixed modes to which we want to fit our model. For the detection of the modes we follow the method given in \cite{2004A&A...428.1039A} applied over each radial order. Rather than applying the detection test to one realisation of the data, we opt to apply the test to a number of rebinned realisations (30 in total, varying from 2 to 60 bins). This amounts to a resolution of $\sim0.03\Delta\nu$ at the high $\nu_{\mathrm{max}}$ values of the stars in our sample. The purpose of including the rebinned realisations is to try and accentuate the features of narrow modes as opposed to noise spikes. A false alarm probability of 10\% is used over a window in frequency of $\Delta\nu/2$, under the assumption of having $\Delta\nu/\left(2N\nu_{\mathrm{bw}}\right)$ independent bins (where $N$ is the number of bins binned over and $\nu_{\mathrm{bw}}$ is the frequency resolution of the rebinned realisation). The detection test is performed for each rebinned realisation and the frequencies of each realisation that exceed the detection threshold are kept. These frequencies are then clustered using the mean shift clustering algorithm \citep{Comaniciu:2002:MSR:513073.513076}\footnote{As implemented in the \texttt{python} package \texttt{scikit-learn} \citep{scikit-learn}.}. We set the minimum number of points needed to constitute a cluster to 5, ensuring that single noise spikes in a few realisations are not kept. The value of 5 was chosen because we empirical found that it was a good trade-off between detecting narrow modes with lower signal-to-noise ratios and detecting spikes due to noise. The identified cluster centres are taken as initial guesses for the mode frequencies in the fitting process (see section~\ref{sec:inc}). In addition, initial guesses for the rotational splitting and inclination angle can also be approximated from the cluster centres. It is important to note that we only search for modes between the $\ell=0$ and $\ell=2$ mode of the same order to avoid ambiguity and issues in the fitting procedure. This is why, in Fig.~\ref{fig:detection}, the mixed mode at just above 245 $\mu$Hz is not used. An example of the mixed modes detected by this algorithm is shown in Fig.~\ref{fig:detection}.

\subsection{Mode Identification}\label{sec:ID}

The detected peaks are of no use by themselves and in order to fit our model to the data we need to know which peaks belong to which mixed mode and whether they are rotationally split. The initial identification of the radial modes is performed using the universal pattern described by \cite{2011A&A...525L...9M}, which is dependent only upon the large frequency separation, $\Delta\nu$.

We follow the formulation for $\ell=1$ mixed modes frequencies given in \cite{2012A&A...540A.143M}

\begin{figure*}
\centering
\includegraphics[width=\textwidth]{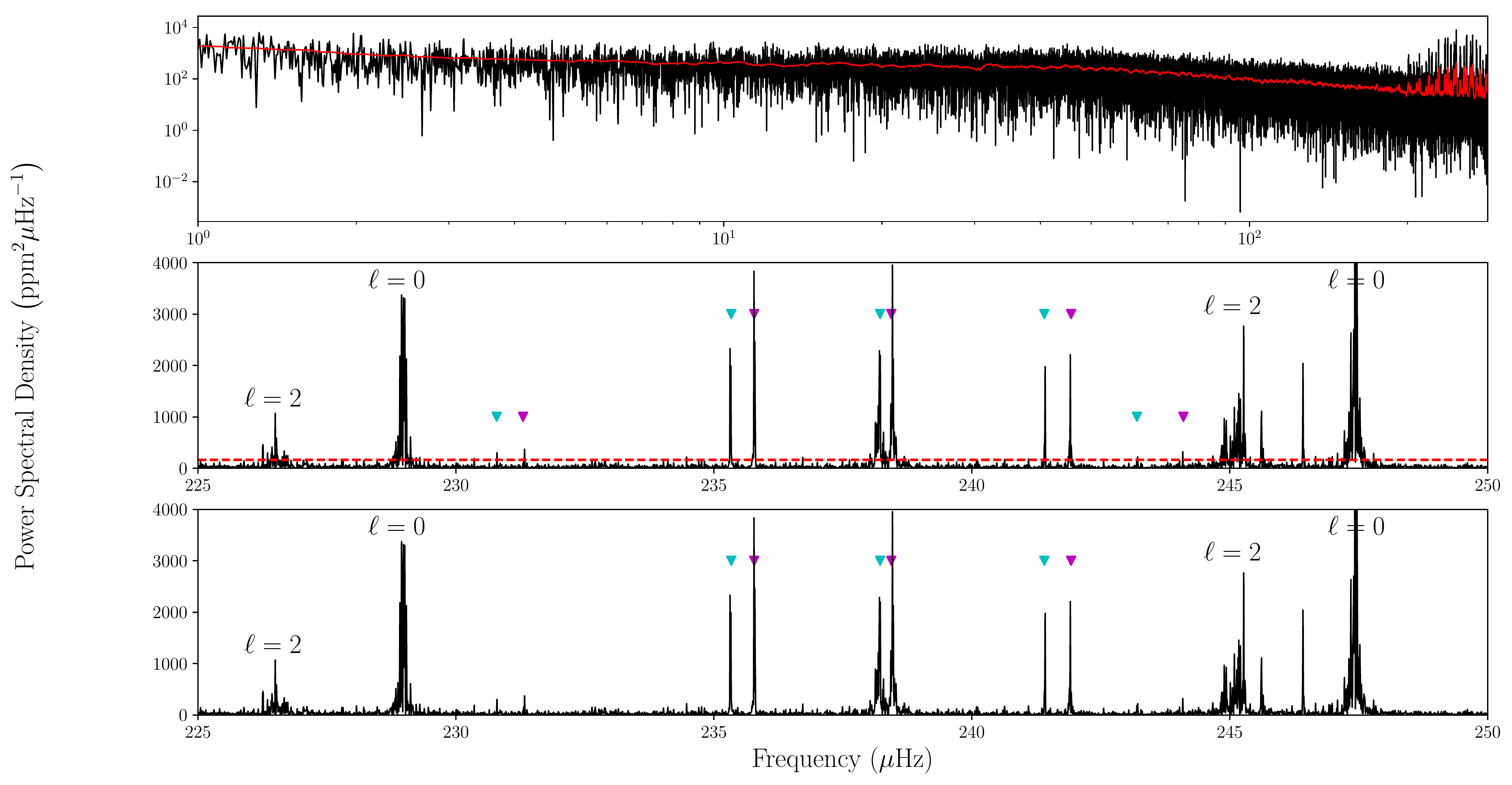}
\caption{The top panel shows a log-log plot of the power spectrum for KIC 5553307 in black and a smoothed spectrum ($1\;\mu$Hz boxcar) in red. The centre and bottom panels show an example radial order taken from the same star. The centre panel shows the modes that have been detected following the method given in section~\ref{sec:det}. The cyan and magenta triangles denote the possible positions of the detected rotationally split components ($m=-1$ and $m=+1$ respectively). Only the $m=\pm1$ components are seen due to the inclination angle of the star being close to 90\degrees. The red dashed line in the centre panel shows the detection threshold assuming a false-alarm probability of 10\%. The bottom panel shows the modes chosen once the additional constraint of the mode identification (see Section~\ref{sec:ID}) has been applied. For example, the peaks at $\sim$230$\mu$Hz, whilst detected, were not used in the subsequent analysis as they do not agree with the assumed mixed mode pattern.}
\label{fig:detection}
\end{figure*}

\begin{equation}
\nu  = \nu_{n_{p},\ell=1} +  \frac{\Delta\nu}{\pi}\arctan\left[q\tan\pi\left(\frac{1}{\Delta\Pi_{1}\nu} - \varepsilon_{g}\right)\right],
\label{eqn:asymp}
\end{equation}
where $\nu_{n_{p},\ell=1}$ is the nominal p-mode frequency, $\Delta\nu$ is the large frequency separation, $q$ is the coupling between the p- and g-modes, $\Delta\Pi_{1}$ is the $\ell=1$ period spacing and $\varepsilon_{g}$ is a phase term. It is assumed that $\varepsilon_{g}=0$, in accordance with \cite{2011Natur.471..608B} and \cite{2012A&A...540A.143M}. This assumption has since been shown to be questionable by \cite{2018A&A...610A..80H}. Given that the parameters were used only to guide the eye this is nevertheless justifiable, since the procedure was more than ample to allow robust identification of the modes. 

Eq.~\ref{eqn:asymp} is combined with the formulation for the expected rotational splittings of $\ell=1$ mixed modes as described in \cite{2013A&A...549A..75G}

\begin{equation}
\frac{\nu_{\mathrm{s}}}{\nu_{\mathrm{s}, \mathrm{max}}} = \zeta\left(1-2\frac{\left\langle\Omega\right\rangle_{\mathrm{env}}}{\left\langle\Omega\right\rangle_{\mathrm{core}}}\right) + 2 \frac{\left\langle\Omega\right\rangle_{\mathrm{env}}}{\left\langle\Omega\right\rangle_{\mathrm{core}}},
\label{eqn:splittings}
\end{equation}
where $\nu_{\mathrm{s}}$ is the rotational splitting, $\nu_{\mathrm{s}, \mathrm{max}}$ is the maximum splitting, $\zeta$ is the ratio of the mode inertia in the g-mode cavity to that of the entire cavity \citep{2015A&A...580A..96D}, $\left\langle\Omega\right\rangle_{\mathrm{core}}$ is the angular rotational velocity averaged over the core regions and $\left\langle\Omega\right\rangle_{\mathrm{env}}$  is the angular rotational velocity averaged over the envelope. We make the assumption that the contribution from the envelope is very small, i.e. that the envelope is rotating very slowly. In the limit $\left\langle\Omega\right\rangle_{\mathrm{env}} \ll \left\langle\Omega\right\rangle_{\mathrm{core}}$ the ratio $\left\langle\Omega\right\rangle_{\mathrm{env}}/\left\langle\Omega\right\rangle_{\mathrm{core}}$ can be neglected, reducing Eq.~\ref{eqn:splittings} to

\begin{equation}
\nu_{\mathrm{s}} = \zeta\nu_{\mathrm{s}, \mathrm{max}}.
\end{equation}

The individual parameters contained in equations \ref{eqn:asymp} and \ref{eqn:splittings} were manually adjusted to produce a good fit by eye. The values derived are approximate and are only used in the generation of the artificial data; otherwise they are not required in the rest of the inference. In other words, we construct a pattern of the approximate $\ell=1$ mixed mode central frequencies and rotational splittings to identify the mixed modes. The mixed mode is selected if we observe the prerequisite number of peaks: one for angles close to zero, two for those close to 90\degrees and 3 for intermediate angles. These provide initial guesses for the central frequency and the rotational splitting of the mixed modes, which are then subsequently fed into the fitting procedure. An example of the final modes selected is given in Fig.~\ref{fig:detection}. It might be expected that by only fitting a subset of our data (due to not detected all components of a mode) that we could introduce a selection effect. However, an underlying assumption that is made during the subsequent analysis is that each mode possesses the same underlying  inclination angle. As a result, this selection effect will only impact the uncertainty of the determination of the inclination angle rather than the underlying value itself.

\subsection{Peakbagging Model}\label{sec:inc}

The determination of the inclination angle uses the formulation described in \cite{2003ApJ...589.1009G}. This relies on the assumption that there is equipartition of energy between mode components of differing azimuthal order in any given multiplet. For $\ell=1$ modes the inclination angle can be derived from the amplitudes of the $|m|=\ell$ and $m=0$ components.

The model that is fitted to each mixed mode is given by (e.g. \citealt{refId0}):
\begin{equation}
\mathcal{M}(\nu; \mathbf{\theta}) = \sum^{\ell}_{m=-\ell} \frac{\mathcal{E}_{1m}(i)H}{1 + \frac{4}{\Gamma^{2}}(\nu-\nu_{0}-m\nu_{s})^{2}}+ B.
\label{model}
\end{equation}
where $B$ is the background (the very narrow region in frequency in which the mixed mode is being fitted results in the use of a flat background being valid), $\nu_{0}$ is the central frequency of the mixed mode (i.e. the frequency of the $m=0$ component), $\nu_{s}$ is the rotational splitting and the summation is over each $m$ component which runs from $-\ell$ to $+\ell$. The height $H$ can be parametrised in terms of the amplitude of the mode, $A$, and the linewidth, $\Gamma$, to alleviate the unwanted impact of the strong anti-correlation between $H$ and $\Gamma$ \citep{2003A&A...398..305C, 2006MNRAS.371..935F}
\begin{equation}
    H = \frac{2A^{2}T}{\pi\Gamma T + 2},
\end{equation}
where $T$ is the length of the observations. This expression is used in order to account for the change in the mode profile when the linewidth tends towards the unresolved regime, i.e. $\Gamma \lesssim 2\nu_{\mathrm{bw}}/\pi$.

The final parameter to introduce is the visibility factor $\mathcal{E}$ \citep{2003ApJ...589.1009G} which is given by
\begin{equation}
\mathcal{E}_{\ell m}(i) = \frac{(\ell - |m|)!}{(\ell+|m|)!}P^{|m|}_{\ell}(\cos i)^{2},
\label{vis}
\end{equation}
subject to the constraint
\begin{equation}
\sum_{m}\mathcal{E}_{\ell m}(i) = 1,
\end{equation}
where $P^{|m|}_{\ell}(\cos i)^{2}$ are associated Legendre polynomials. For the case of $\ell=1$ modes, Eq~\ref{vis} gives
\begin{align}
\mathcal{E}_{1,0}(i) &= \cos^{2}(i),\\
\mathcal{E}_{1,\pm1}(i) &= \frac{1}{2}\sin^{2}(i).
\label{eqn: inc_leg}
\end{align}

\begin{figure*}
\centering
\subfloat{\includegraphics[width=0.8\textwidth]{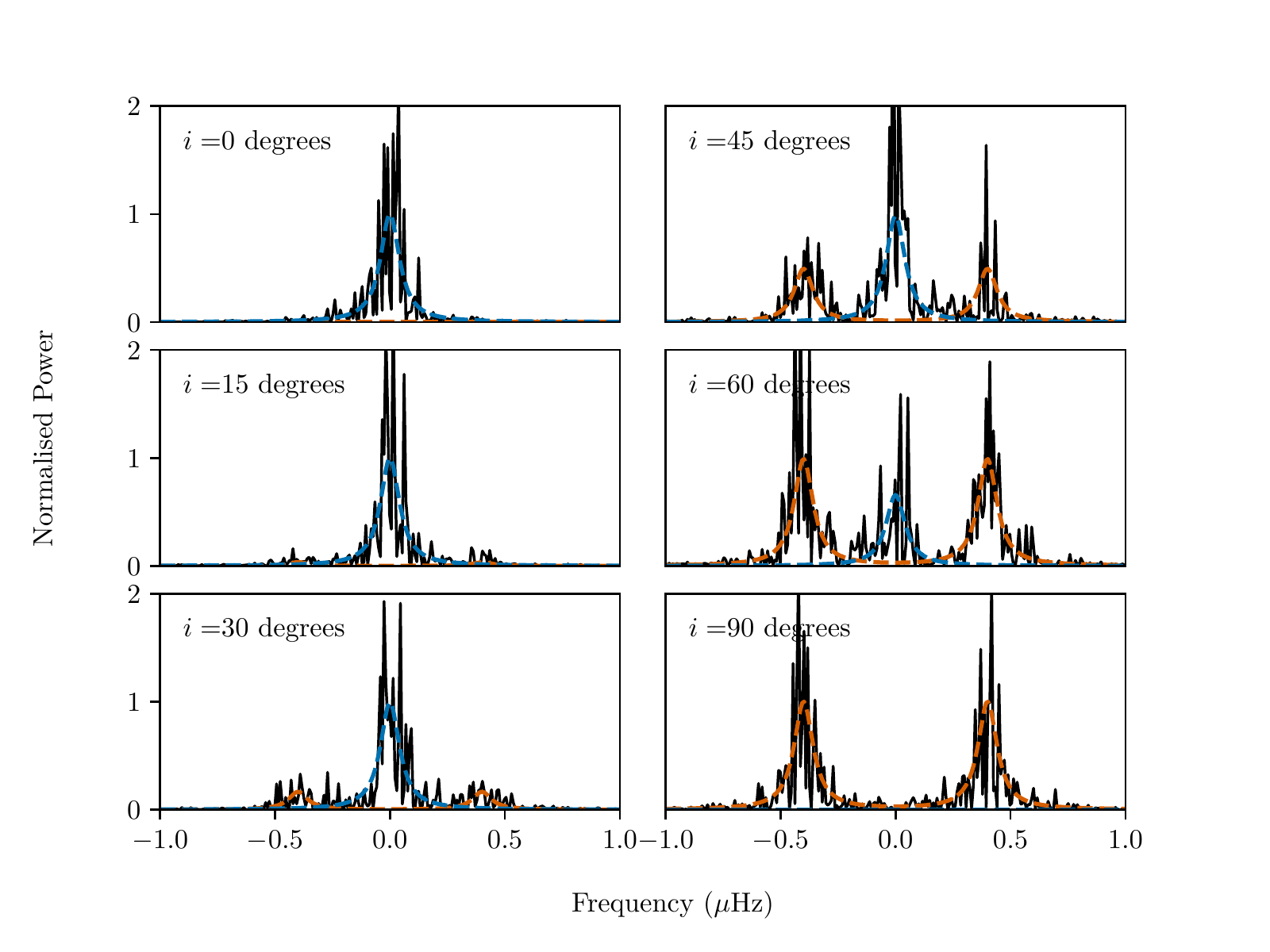}}
\caption{Example $\ell=1$ modes observed at different inclination angles. The $m=\pm1$ components are shown in orange and the central $m=0$ component is shown in blue. The angle of inclination used in simulating the mode is labelled in each panel. Furthermore, a mode linewidth of $0.2\mu$Hz, a rotational splitting of $0.4\mu$Hz and an observing length of 4 years were used in all cases.}
\label{fig: splittings_nice}
\end{figure*}

The mode visibility therefore modulates the amplitudes of the modes in a way that is dependent upon the inclination angle of the star. If we were to observe a star at 90\degrees (i.e. equator-on) with respect to our line of sight then in the power spectrum we would only see the outer components of the $\ell=1$ mixed modes (i.e. $m=\pm1$). Whereas if we were to observe a star pole-on, at close to 0\degrees with respect to our line of sight, then in the power spectrum we would only observe the central ($m=0$) component. Intermediate angles would fall in between the two extremes in terms of relative amplitudes. An example of how the inclination angle modulates the component amplitudes of an $\ell=1$ mode is shown in Fig.~\ref{fig: splittings_nice}.

\subsection{The Fitting Process}\label{sec:fit}

Our interest lies in fitting individual rotationally split $\ell=1$ mixed modes rather than performing a global fit. There are many mixed modes per order and the high quality \emph{Kepler} data enable us to extract many individual measures of the inclination angle.

In this work we adopt a Bayesian framework for fitting the mixed modes (see e.g \citealt{refId0,2014A&A...570A..41K,2016MNRAS.456.2183D}) since we want to obtain posterior probability distributions for the inclination angle to be used in the subsequent analysis. We use Markov Chain Monte Carlo (MCMC) to sample the parameter space, making use of \texttt{ptemcee}\footnote{\texttt{ptemcee} is a parallel tempering extension to \texttt{emcee} that uses dynamic temperature selection.} \citep{2016MNRAS.455.1919V, 2013PASP..125..306F}. In subsequent sections we will adopt this procedure as well.

The log-likelihood function used in the model fitting assumes the noise properties of the power spectrum follows a $\chi^{2}_{2}$ distribution and is given by \citep{1986ASIC..169..105D, 1990ApJ...364..699A}
\begin{equation}
\ln L = -\sum_{\nu}\left[\ln M_{\nu}(\mathbf{\theta}) + \frac{P_{\nu}}{M_{\nu}(\mathbf{\theta})}\right],
\end{equation}
where the summation is made over each frequency bin, $M_{\nu}(\mathbf{\theta})$ is the model evaluated at a given frequency for a set of parameters $\mathbf{\theta}$, and $P_{\nu}$ is the power at a given frequency.

The prior distributions were taken to be uniform for all parameters apart from the central frequency of the mode and the inclination angle. A Gaussian prior was placed on the central frequency with a mean according to the approximate mode frequency extracted from the clustering (see section~\ref{sec:det}) and a standard deviation of 0.2 $\mu$Hz. This is a weakly informative prior, the standard deviation of which is taken to be a value that is narrow enough that the central frequency is fitted to the central component of the mixed mode and wide enough to allow for the fact that the initial guess may not exactly coincide with the underlying value. Finally an isotropic prior, $p(i) \propto \sin i$, was placed on the inclination angle which comes from the assumption that stars are oriented randomly with respect to the observer. The isotropic prior is also uninformative and preferable to a uniform prior, as shown in more detail in Appendix~\ref{sec: why_uninform}. The prior distributions used are given in Table~\ref{tab:priors}.

\begin{table}
	\centering
	\caption{Model prior distributions for the peak-bagging analysis. $\mathcal{N}$(mean,standard deviation) indicates Gaussian priors, and $\mathcal{U}$(lower bound, upper bound) indicates uniform priors.}
	\label{tab:priors}
	\begin{tabular}{ll} 
	\hline
	Parameter & Prior\\
	\hline
	$A$ & $\mathcal{U}(0,50)$ (ppm)\\
	$\nu_{0}$ &$\mathcal{N}(\nu_{0,\mathrm{initial}},0.2)$ ($\mu$Hz)\\
	$\Gamma$ & $\mathcal{U}(0,5)$ ($\mu$Hz)\\
	$\nu_{\mathrm{s}}$ & $\mathcal{U}(0, 0.7)$ ($\mu$Hz)\\
	$i$ & $\sin i$ (degrees)\\
	$B$ & $\mathcal{U}(0, 200)$ (ppm$^{2}\mu$Hz$^{-1}$)\\
	\hline
	\end{tabular}
\end{table} 

\begin{figure}
\centering
\subfloat{\includegraphics[width=0.45\textwidth]{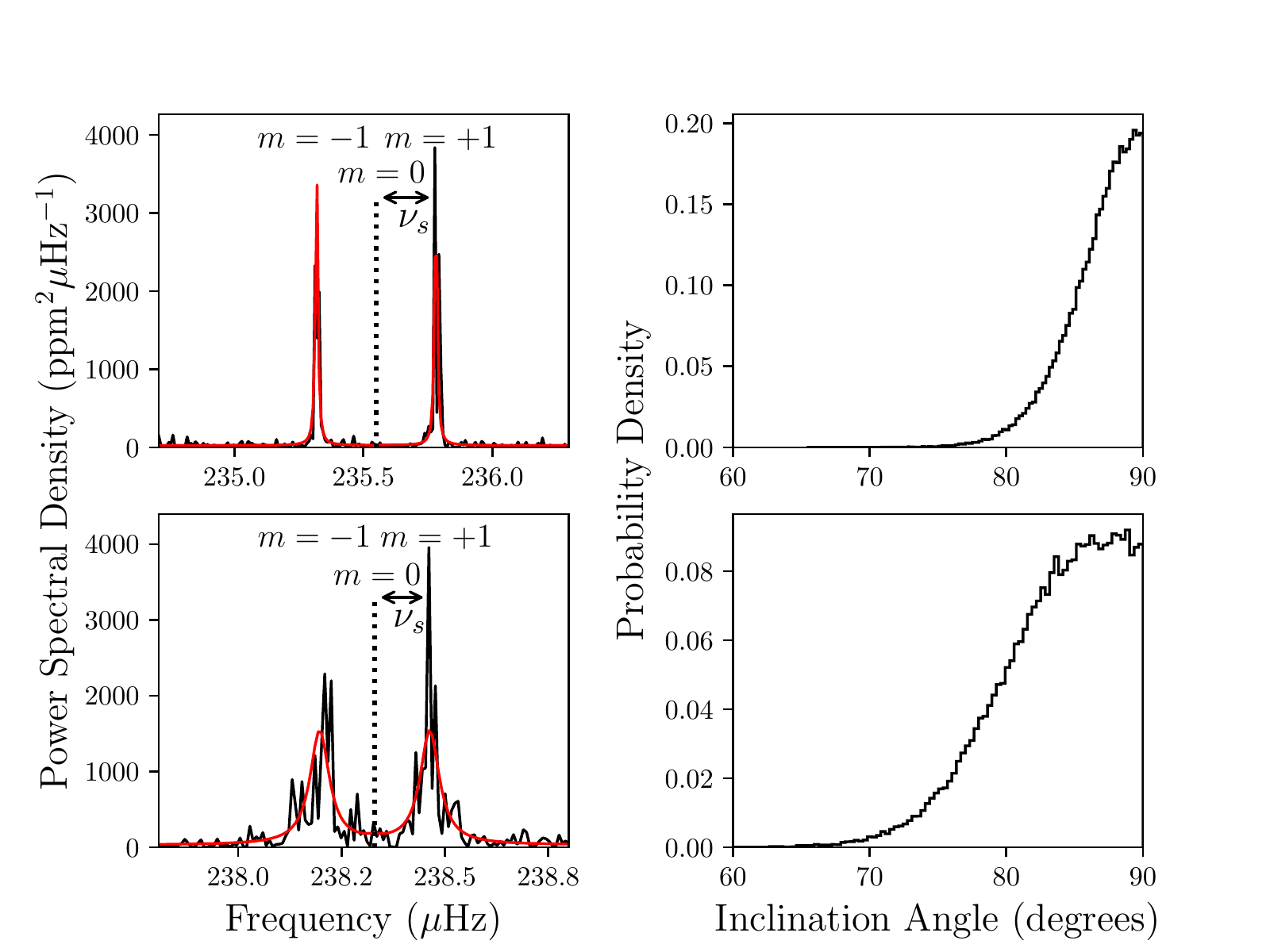}}
\caption{An example of two Lorentzian fits made to modes in KIC 5553307 are shown in the left-hand side and their respective inclination angle posterior probability distributions are shown on the right-hand side. The top panel shows a more g-dominated $\ell=1$ mixed mode at $235.5\;\mu$Hz, and the resolved p-dominated $\ell=1$ mixed mode at $238.3\;\mu$Hz is shown in the bottom panel. On the left-hand side, the power spectrum is shown in black with the best fit model overplotted in red. The respective $m$ components are also labelled as well as a demonstration of the rotational splitting $\nu_{s}$. The black dotted line denotes the estimated position of the $m=0$ component which is not visible due to the close to 90\degrees angle of inclination of the star.}
\label{fig: splittings}
\end{figure}

This approach yields posterior distributions of the inclination angle for individual modes rather than point estimates, which will be useful in the following sections. An example of fits made to both p- and g-dominated modes is shown in Fig.~\ref{fig: splittings}.

\section[]{Extracting the inclination angle}\label{sec: method1}

\subsection{Hierarchical Method}

We adopt an approach in this work that explicitly models the underlying distribution whilst accounting for the initial prior distribution. This is performed following the hierarchical Bayesian method described in \cite{2010ApJ...725.2166H} and again using MCMC (see section~\ref{sec:fit}) to sample the parameter space. The explicit modelling of the distribution should enable us to better approximate its underlying shape, especially in the case of truncation at low and high inclination angles. The method is hierarchical because we are allowing the prior distribution to be modelled. We are essentially trying to find a model that better describes the data than the original uninformative prior. This enables us to infer the underlying inclination angle from the model given the posterior samples that we gathered in the previous section. 

In order to infer the underlying angle, we turn to the marginalised likelihood given in \cite{2010ApJ...725.2166H} (see Appendix~\ref{sec:like_deriv} for a full derivation)

\begin{equation}
\mathcal{L_{\alpha}} \approx \prod^{N}_{n=1}\frac{1}{K}\sum^{K}_{k=1}\frac{f_{\alpha}(i_{nk})}{p_{0}(i_{nk})}.
\label{eqn: new_like}
\end{equation}
where $f_{\alpha}(i_{nk})$ is the model we want to fit with parameters $\alpha$, evaluated at the $k$th sample of the posterior of the $n$th mode. The parameter $p_{0}(i_{nk})$ is the uninformative prior (used in the original fitting process, as denoted by the subscript 0) evaluated for the same sample. 

In addition, Eq.~\ref{eqn: new_like} can be interpreted as follows: inside the sum is the ratio of the new prior to the initial (uninformative prior), taken over each mode. The prior used in the initial fitting process should be uninformative and be valid over the entire range occupied by the new prior, otherwise the inferred parameters can be very uncertain. Eq.~\ref{eqn: new_like} is a marginalised likelihood because we have integrated out all of the parameters of the original fit. We have therefore assumed that the distribution over the parameters from the original fit is separable, which has enabled us to formulate the likelihood function in terms of just the inclination angle. By assuming separability, we have also assumed that the only distribution that needs to be adjusted from the assumed prior to match the population distribution is the inclination angle.

We adopt the hierarchical method because it allows us to combine the multiple observations of inclination angle (one per mixed mode) in a principled way, accounting for the uncertainty in each observation and also possible systematics in the amplitudes and rotational splitting of each mode. In order to model the systematics, we do not assume that the inclination parameters measured by the fit to each mode are identical, we rather assume that they come from a distribution with a central peak and scatter; see Section~\ref{sec:model}. By fitting for both the central value and the scatter (i.e. the location and concentration parameters defined in Section~\ref{sec:model}), we can obtain an estimate of the inclination angle and the degree of systematic scatter in the mode-to-mode values simultaneously, all while accounting for the uncertainties in each mode's estimate of the inclination angle.

\subsection[]{Inclination angle model of the hierarchical method}\label{sec:model}

The model we adopt for $f_{\alpha}(i)$ is a slightly modified version of the Fisher distribution (as chosen in \citealt{2009ApJ...696.1230F}) that is equivalent to a zero-mean Normal distribution on the sky. Whilst the original formulation of the Fisher distribution is a valid model if we assume our angle of inclination distribution is (or is very close to) isotropic, it is unsuitable for the highly localised inclination angle distributions of individual stars. 

To add flexibility to the model, a location parameter is added that enables the peak of the distribution to be shifted in the region $\mu\in[0, \pi/2]$. This results in the model being able to adapt to both a sharply peaked distribution (in the case of the individual stars) and a much wider population distribution. It offers the ability to model an isotropic distribution and one localised towards any angle between 0 and 90 degrees (i.e. towards anisotropy). The location parameter will provide information on the underlying inclination angle of the star and so this is our parameter of interest. The updated form is given by
\begin{equation}
f(i|\mu,\kappa) = \exp\left[\kappa\cos\left(i - \mu\right) \right]\sin i,
\label{eqn: model}
\end{equation}
where $\kappa$ is the concentration parameter and the additional parameter $\mu$ is the aforementioned location parameter. We have used $f$ to explicitly state that the function is not a probability distribution, and will proceed to use $p$ otherwise. In its current state the distribution in Eq.~\ref{eqn: model} is unnormalised.  The normalisation constant can be derived analytically (as shown in Appendix~\ref{sec: norm_deriv}) to give the probability distribution
\begin{equation}
p(i|\mu,\kappa) = \left\{I_{0}(\kappa) + 2\varphi(\kappa,\mu)\right\}^{-1}
\exp\left[\kappa(i-\mu)\right]\sin i,
\end{equation}
where $I_{0}(\kappa)$ is a modified Bessel function of the first kind of order zero, and $\varphi(\kappa,\mu)$ is a function of the Fisher distribution parameters used in the normalisation of the distribution (for more information see Appendix~\ref{sec: norm_deriv}).

\begin{figure}
\centering
\subfloat[]{\includegraphics[width=0.45\textwidth]{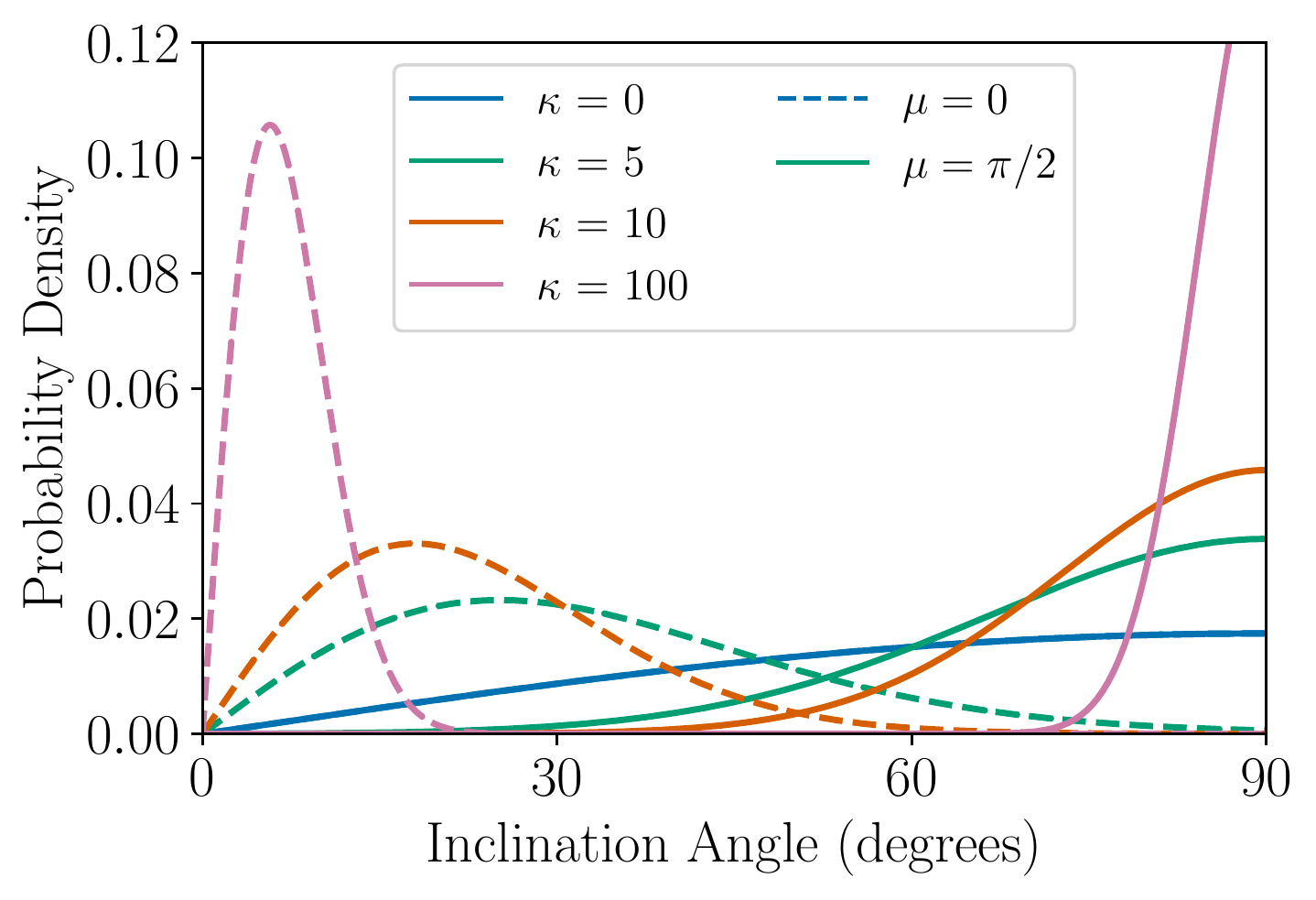}}
\vfill
\subfloat[]{\includegraphics[width=0.45\textwidth]{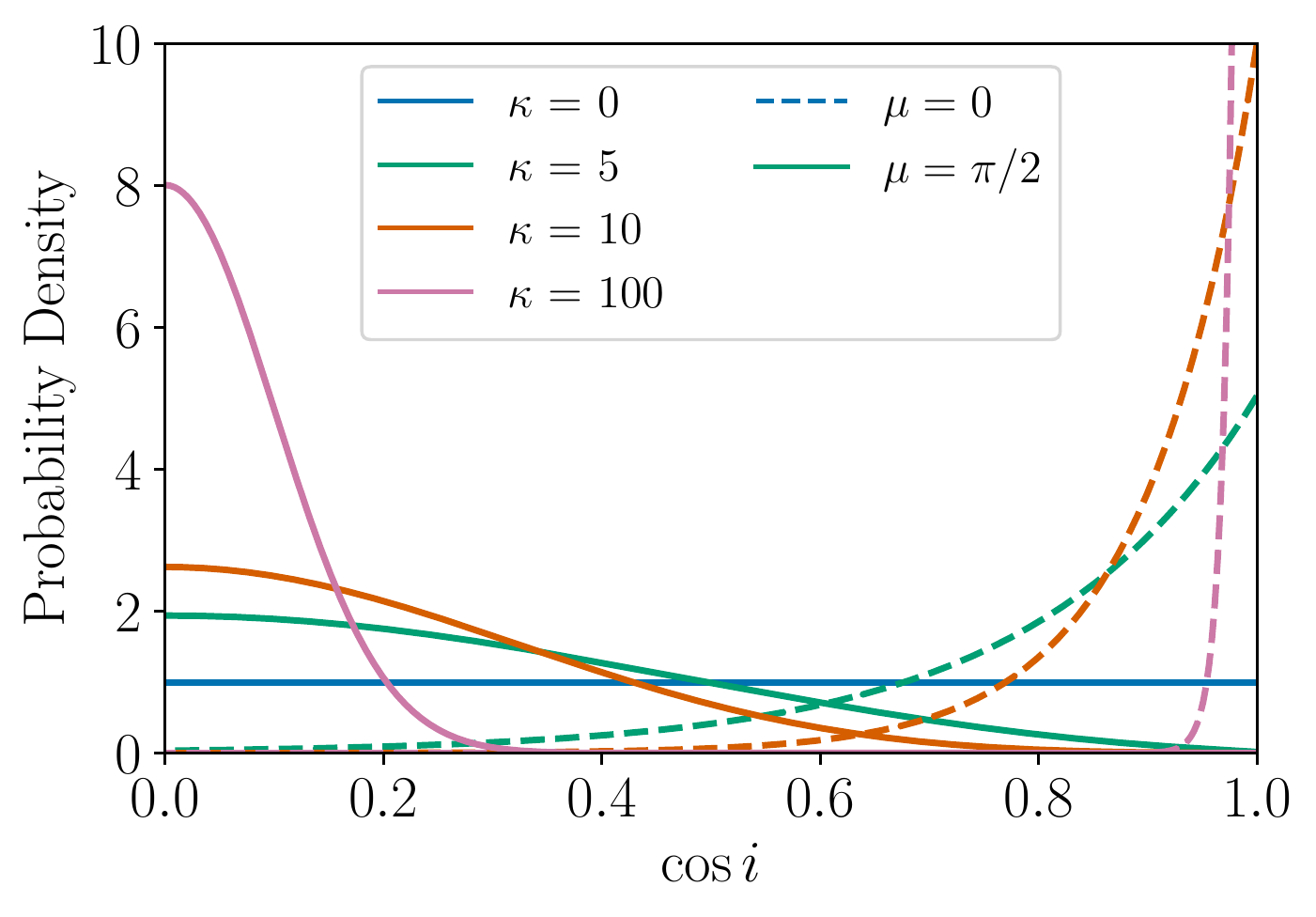}}
\vfill
\subfloat[]{\includegraphics[width=0.45\textwidth]{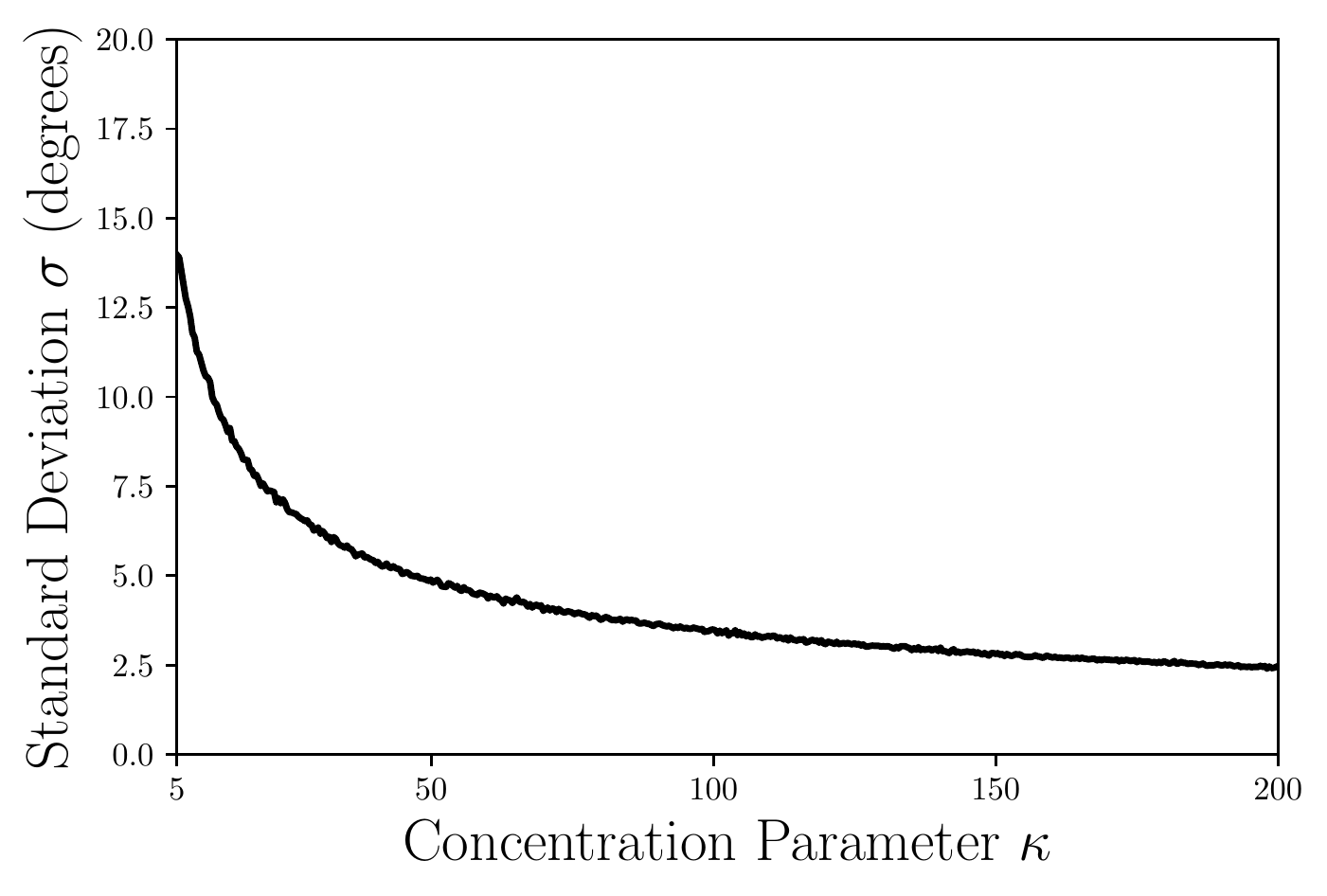}}

\caption{A demonstration of the model used in the hierarchical modelling as a function of $i$ and $\cos i$. The top panel shows the modified Fisher distribution as a function of $i$ for a variety of different $\kappa$ values. The solid lines denote distributions computed with $\mu=\pi/2$ and the dashed lines show distributions computed with $\mu=0$. The middle panel shows the distributions in the top panel as a function of $\cos i$ using equation \ref{new_fisher}. The bottom panel shows how the parameter $\kappa$ affects the standard deviation of the distribution, with larger $\kappa$ leading to a narrower distribution.}
\label{fig: new_fisher}
\end{figure}

Examples of this modified Fisher distribution, which will become our model in the hierarchical method, are shown in the top panel of Fig.~\ref{fig: new_fisher}. The distribution can represent isotropy when $\kappa=0$, and the addition of the location parameter also allows for deviations from isotropy to be accounted for. The average measured inclination angle under the assumption of isotropy is $\sim57.2$ degrees (1 radian) and so the expected value of $\mu$ should also take the same value under isotropy\footnote{The location parameter $\mu$ will be fitted in radians and so will be referred to and displayed as such, otherwise we will refer to the angles in degrees.}. Therefore, subtle deviations or possible biases could be inferred from this parameter and not just from $\kappa$. In addition, the standard deviation of the distribution as a function of $\kappa$ is given in the last panel of Fig.~\ref{fig: new_fisher}. This helps show how the variance of the distribution decreases with increasing $\kappa$ leading to a much more localised and sharply peaked distribution.

An important consideration in the hierarchical analysis is the support (i.e. region of validity in parameter space) of both the initial prior and the new prior model in Equation~\ref{eqn: new_like}. It is advisable to avoid priors that tend to zero at the edges of the parameter space since this can cause problems in the likelihood function (Eq.~\ref{eqn: new_like}), due to the division. This occurs if the model is fitted in inclination angle space due to the isotropic prior tending to zero with the inclination angle. However, this effect can be mitigated by instead performing the inference in $\cos i$ whereby the isotropic prior becomes uniform (see Appendix~\ref{sec: why_uninform}). Therefore the modified Fisher distribution as a function of $i$ (as given above) is not adequate in the modelling and so must be transformed such that we obtain the distribution in $\cos i$, given that $i$ is Fisher distributed with some concentration parameter $\kappa$ and location parameter $\mu$. 

Following the method given in \cite{2014ApJ...796...47M} for transforming the original Fisher Distribution, we can transform this updated model (as shown in Appendix~\ref{sec: model}) to derive the following probability distribution

\begin{equation}
p(y | \mu,\kappa) = \left\{I_{0}(\kappa) + 2\varphi(\kappa,\mu)\right\}^{-1}
\exp\left(\kappa y\cos\mu + \kappa\sqrt{1-y^{2}}\sin\mu\right),
\label{new_fisher}
\end{equation}
where $y = \cos i$.

An isotropic prior was imposed on the location parameter $\mu$ in order to reflect the isotropic prior distribution on the angle during the fitting process. The application of an uninformative prior on $\mu$ is important, as we shall see in the next section. When inferring the inclination angle distribution for individual stars, the distributions are expected to be much narrower than the population distribution (since we are searching for localised distributions). As a result, large values of $\kappa$ are to be expected and the prior adopted by \cite{2014ApJ...796...47M},  $p(\kappa) = (1 + \kappa^{2})^{-3/4}$, is too aggressive and prevents large $\kappa$ values from being explored. Therefore, we instead adopt a Half-Cauchy prior with a width, $\gamma=50$, on $\kappa$ (e.g. \citealt{2009arXiv0901.4011G}):

\begin{equation}
p(\kappa | \gamma) = \frac{1}{\pi\gamma}\left(\frac{\gamma^{2}}{\kappa^{2} + \gamma^{2}}\right)
\;\text{ for}\;\kappa>0.
\end{equation}
The extended tail of the Half-Cauchy distribution enables much larger $\kappa$ values to be explored if necessary. The width of the Half-Cauchy prior was tested using values from 10 to 100 and in all cases consistent values were returned indicating that the choice of $\gamma$ is not adversely affecting the inferred parameters.

\section{Application to data}

\subsection{Artificial data}

The method was first tested on artificial red giant power spectra with properties drawn from the same parameter space as the real data and with inclination angles drawn from a known distribution (i.e. isotropic). This testing shows the expected efficiency when applied to real data, such that the accuracy and precision of the method can be assessed. Where available, the same parameters are used in the construction of these power spectra (e.g. estimated period spacings, coupling factors etc.). The construction of the artificial datasets is discussed in Appendix~\ref{sec: artif_spec}\footnote{The code used to generate these power spectra can be found at \url{https://github.com/jsk389/Artificial-red-giant-power-spectra}.}.

The mixed modes in the power spectrum are identified and fitted in the manner described in Section~\ref{sec:analysis}, resulting in inclination angle posterior distributions of every fitted mode for each star in the sample. A total of 77 of the 90 stars in the simulated sample rotationally split modes were fitted properly. This is because the rotational splitting values are drawn from the estimated rotational splittings extracted whilst performing the mode identification on the real stars. Since there are stars for which the rotational splitting could not be observed (either because of an inclination angle of zero or being a very slow rotator) their rotational splitting estimates were set to zero. As a result there was a chance of drawing a non-zero inclination angle from the input distribution and drawing the rotational splitting parameter from a star with no observed rotational splitting. This would result in an inferred inclination angle inconsistent with the input angle.

\begin{figure}
\centering
\subfloat[]{\includegraphics[width=0.45\textwidth]{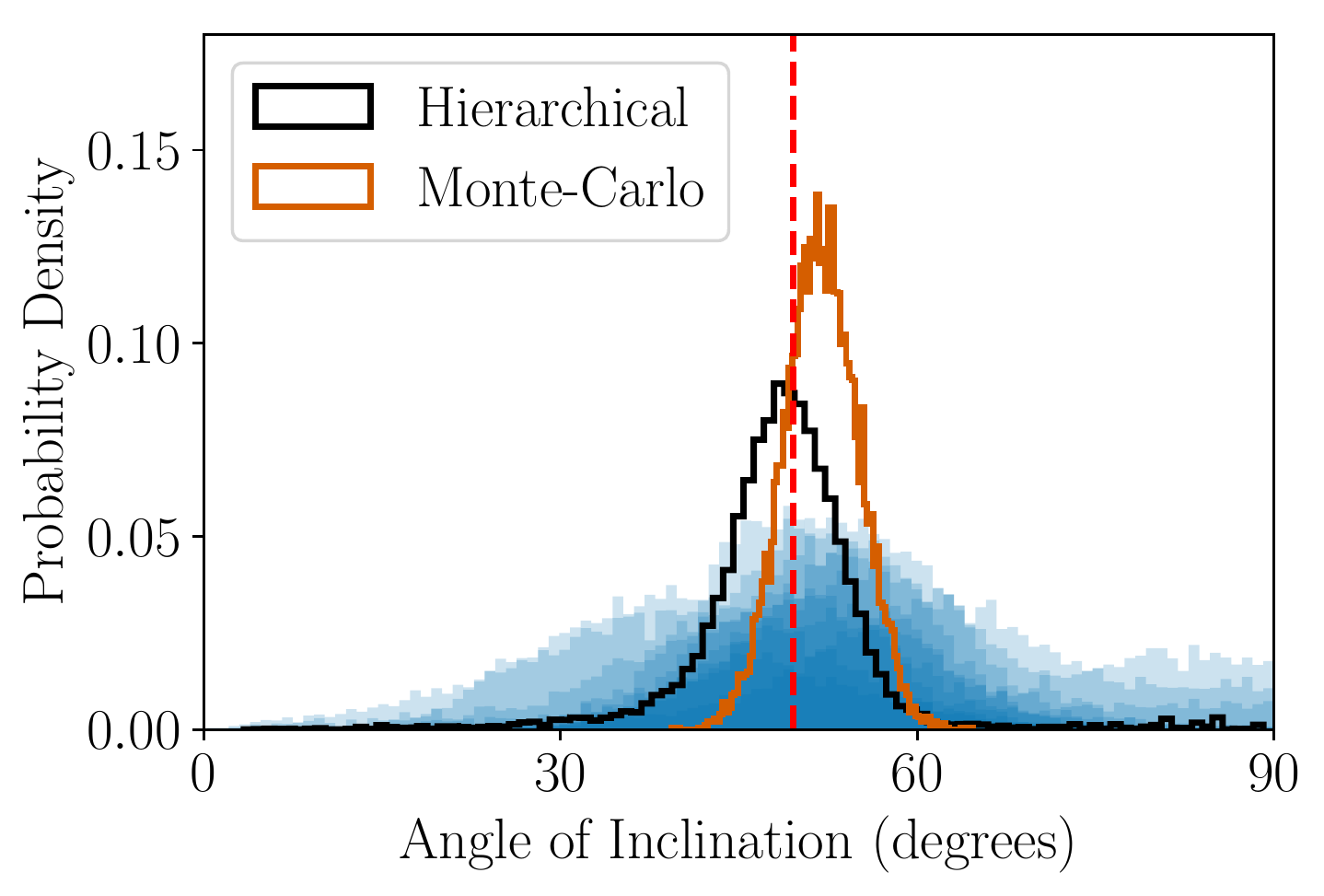}}
\vfill
\subfloat[]{\includegraphics[width=0.45\textwidth]{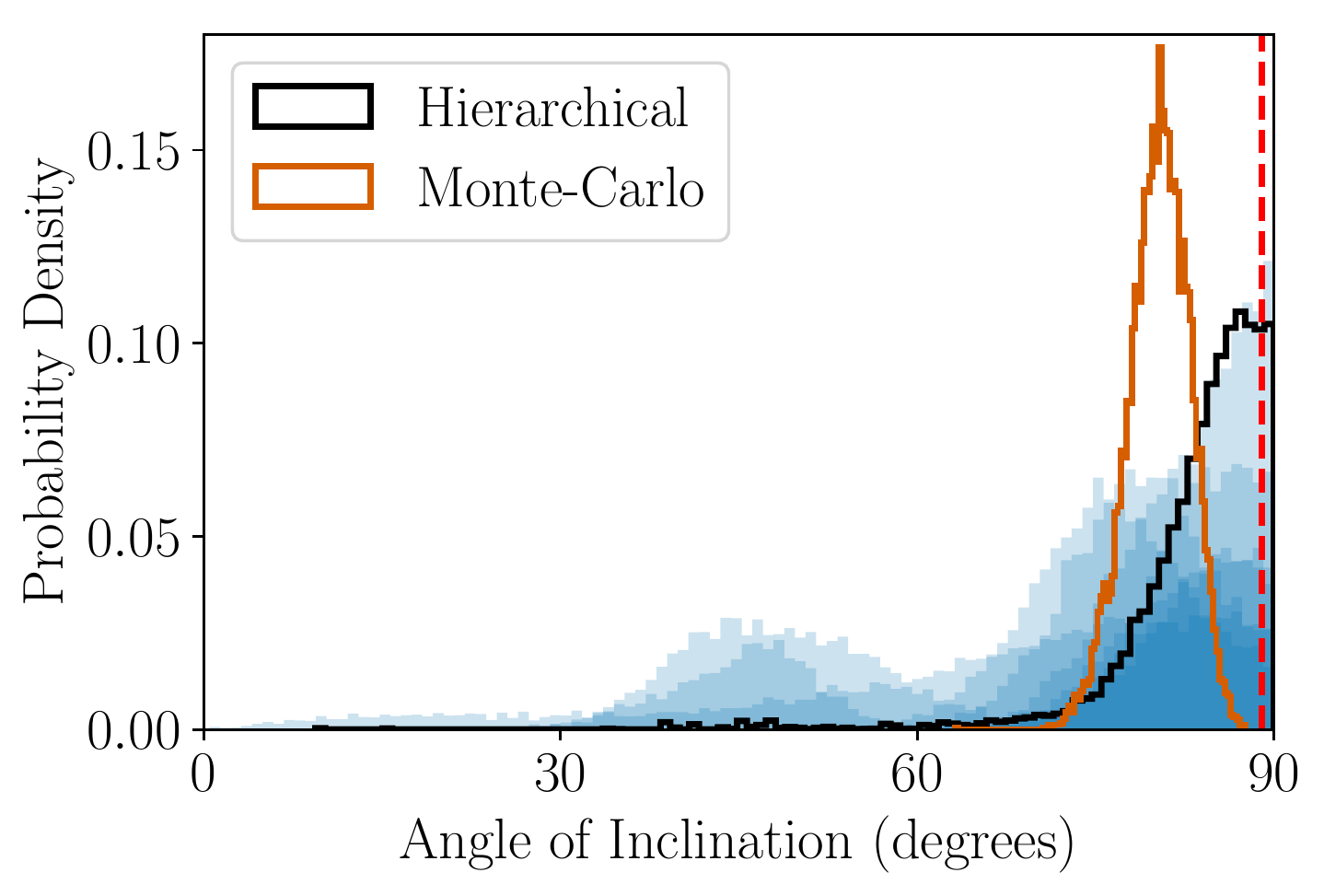}}
\caption{Examples of the fitting procedure applied to two stars from the artificial sample, the top panel shows a star with an input angle (denoted by the red dashed line) of $\sim50^{\circ}$ and the bottom panel shows a star with an input angle close to 90\degrees. The individual overlapping posteriors for each mode are plotted in blue. The mean inclination angle inferred from the Monte-Carlo estimate is shown in orange and the posterior PDF of the location parameter $\mu$ is shown in black.}
\label{fig: artif_example_1}
\end{figure}

Examples of the application of the hierarchical analysis to a few stars in the artificial sample are shown in Fig.~\ref{fig: artif_example_1}. A Monte-Carlo analysis, performed by computing the weighted mean, is also shown for comparison. The weighted mean is computed by drawing one value from the inclination angle distribution of each mode and calculating the mean of the drawn values weighted by the inverse of the variance of the distribution it was drawn from. This is then repeated 1000 times to build up a distribution of the weighted mean. In the top panel, the extended tails towards 90\degrees bias the Monte-Carlo estimate away from the input angle, although this is only slight for an input angle close to 50\degrees (where such tails should not occur often). This highlights the benefit of using the hierarchical inference, since the effect of the prior, which, as discussed earlier, is a major contributor to these tails, is greatly reduced and the posterior PDF of $\mu$ is nicely centred about the input angle. A more extreme case is shown in the bottom panel of Fig.~\ref{fig: artif_example_1} where the input angle is close to 90\degrees and the posterior PDFs of the inclination angle show more structure that is indicative of a lower signal-to-noise fit. Even in such a case, it is reasonably clear as to approximately where the input angle should lie. The Monte-Carlo estimate has been biased by the truncation of the data at 90\degrees.

\begin{figure}
\centering
\includegraphics[width=0.45\textwidth]{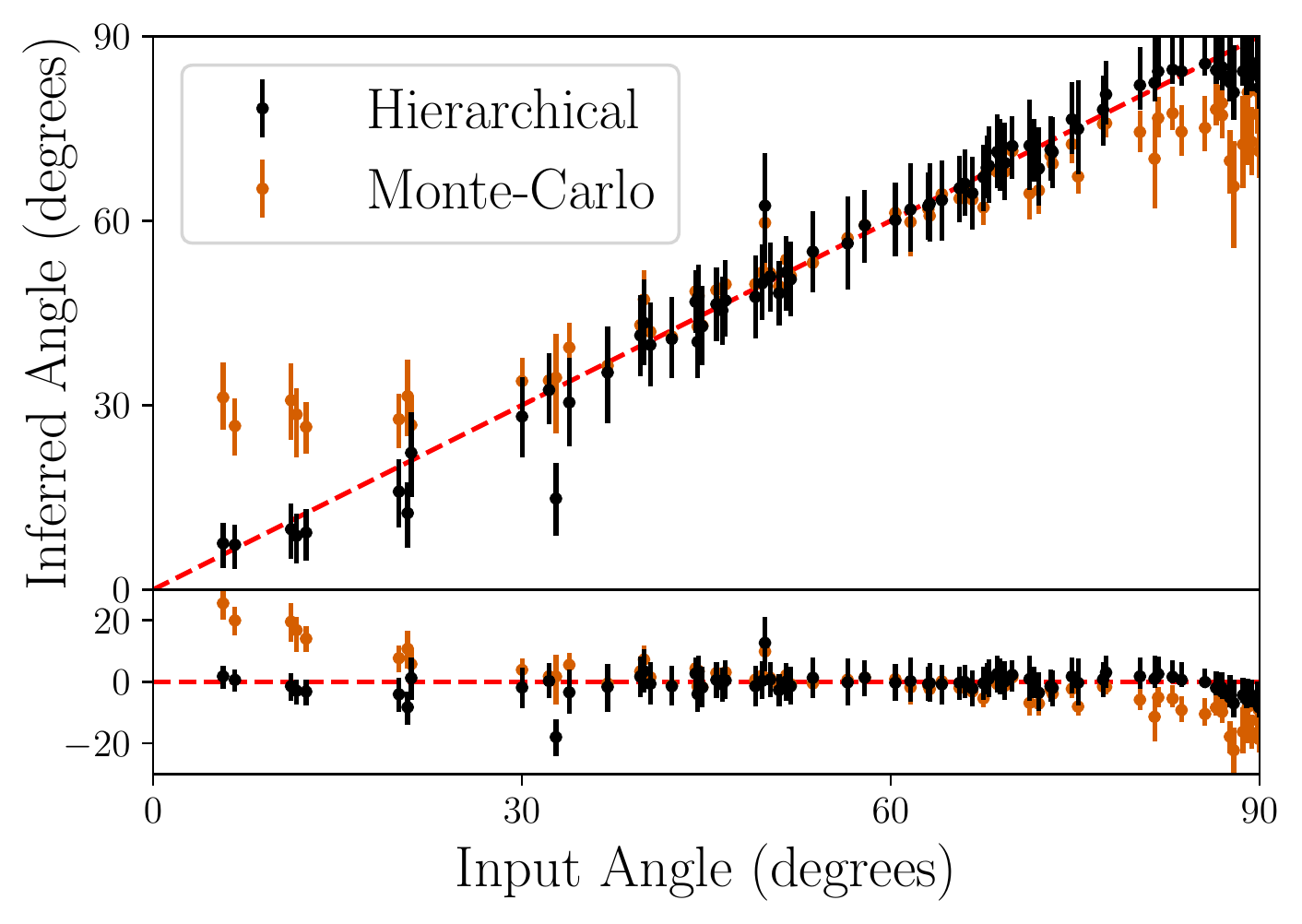}
\caption{The inferred inclination angle shown against the input inclination angle for the artificial data generated with an underlying isotropic distribution. The inferred angles from the hierarchical analyses are shown in black, and those from the Monte-Carlo estimate are shown in orange. In both cases the error bars are taken to be the 68.3\% highest posterior density (HPD) interval. The residuals of the data about the 1:1 line (the red dotted line) are shown at the bottom.}
\label{fig: artif_comparison}
\end{figure}

The results of the hierarchical method applied to the full artificial dataset is shown in Fig.~\ref{fig: artif_comparison}. In the vast majority of cases, the inferred inclination angle (i.e. the posterior distribution of the location parameter) is consistent with the input angle from the posterior estimates. This is promising and shows that we can recover the input angle and in fact all but one of the stars in the sample lies within the 95\% highest posterior density (HPD) credible interval of the underlying value which shows excellent agreement. The results from the hierarchical analysis verify our interpretation of the updated model that the location parameter reflects the underlying angle we wish to infer.

The hierarchical model is not perfect however, and it is important to explain the apparent structure present for inclination angles above 80\degrees and at low angles. Both can be explained by the fact that the parameter $\kappa$ becomes much more important close to the edge of the region of support, i.e. close to 0\degrees or 90\degrees. For large $\kappa$ values ($\kappa > 50$, i.e. a narrow distribution), there is degeneracy between $\kappa$ and $\mu$ when $\mu\rightarrow\pi/2$ and so even a large change in $\kappa$ will not result in a significant change in the shape of the distribution. This uncertainty will propagate through into the posterior PDF of $\mu$. In the 90\degrees case this is evident, whereby the posterior peaks very close to 90\degrees and we cannot say any more precisely as to where the mean value should lie. This idea will be expanded upon in section~\ref{sec:discussion}.

For comparison we also include Monte-Carlo estimates of the weighted mean of the inclination angle posterior distributions, which are shown in Fig.~\ref{fig: artif_comparison}. It is clear that the Monte-Carlo estimate performs poorly close to 0\degrees and 90\degrees as expected due to the effects of the prior and truncation respectively. In the approximate region of 30\degrees to 70\degrees, both the Monte-Carlo estimate and the hierarchical analysis give very similar results. The poorer performance of the Monte-Carlo method is evident by the fact that 22 stars are not within the 95\% HPD credible interval, which occurs at very low ($i<20^{\circ}$) and high ($i > 80^{\circ}$) angles. 

The reasons for the Monte-Carlo estimate failing to accurately calculate the underlying inclination angle across all of parameter space can be summarised as follows. Firstly, when the inclination angle is difficult to constrain, the prior can have a non-negligible contribution to the posterior distribution. This is especially apparent because our prior on the inclination angle is not uniform and a large amount of probability mass is contained close to 90 degrees. As a result the posterior distributions in angle of lower signal to noise modes can have an extended tail towards higher angles. The presence of such a feature would cause a bias in both the mean and variance of the Monte-Carlo estimate. Correcting the effect of the prior \emph{a posteriori} should be avoided and it would be much better to account for this at the same time as approximating the underlying angle. Secondly, the Monte-Carlo estimate fails as a result of the way we measure the inclination angle of the star. Due to the nature of the inclination angle and its symmetries (e.g. \citealt{2003ApJ...589.1009G, 2017NatAs...1E..64C}), we cannot distinguish between an angle $i$ or $(180 - i)^{\circ}$ and so our inclination angles are restricted to lie with the range $i\in [0, 90]^{\circ}$. This introduces the problem of truncation near $i=90^{\circ}$ that will cause the Monte-Carlo estimate to bias low away from $i=90^{\circ}$ due to where the mean lies for the truncated distribution. Additionally this method will bias high when the angle of inclination is close to $i=0^{\circ}$, due to the extended tail seen in the posterior probability density functions (PDFs) of low inclination angle stars. This highlights a need to account for the presence of the prior, which in turn is, to a large extent, responsible for those extended tails. Examples of this can be seen in Fig.~\ref{fig: artif_example_1}.

\subsection{Application to real data}

We applied the same method described above for mode identification, detection and fitting to the \emph{Kepler} red giants in our sample, resulting in posterior PDFs for the inclination angle of each star. The sample dropped from 90 to 89 stars as for one star only a single mode could be fitted and there is no means to check the consistency of the fit. A table of the inferred inclination angles are given in Table~\ref{tab:real_incs}.

\section[]{Population Distribution}

Now that we have obtained inclination angles for individual stars, we can also go one step further and assess whether or not this population of stars is isotropically (randomly) distributed in angle. To do so we opt for a straightforward approach based around the Kolmogorov-Smirnov (K-S) test (see e.g. \citealt{doi:10.1080/01621459.1951.10500769}).

\begin{figure}
\centering
\subfloat[]{\includegraphics[width=0.45\textwidth]{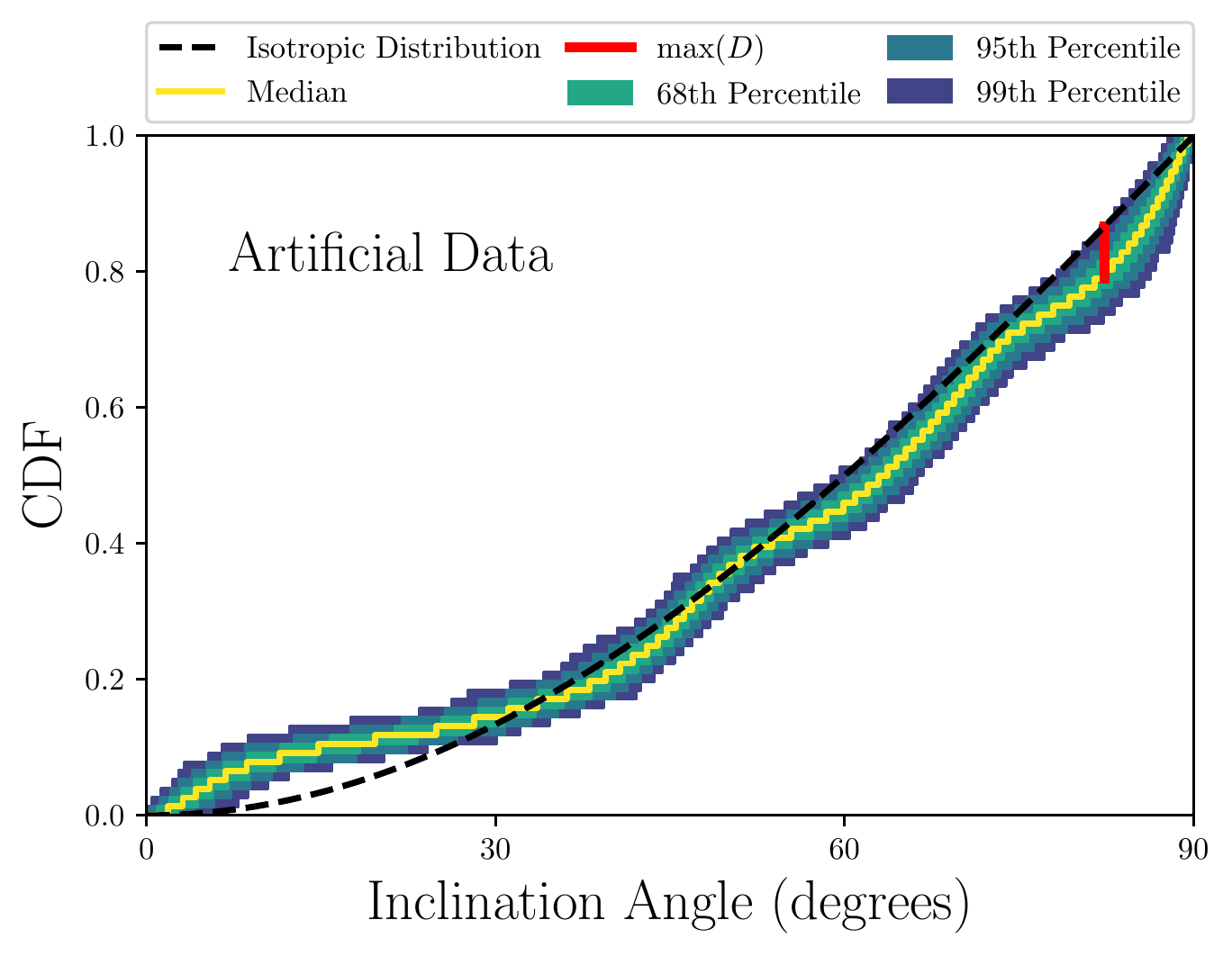}}
\vfill
\subfloat[]{\includegraphics[width=0.45\textwidth]{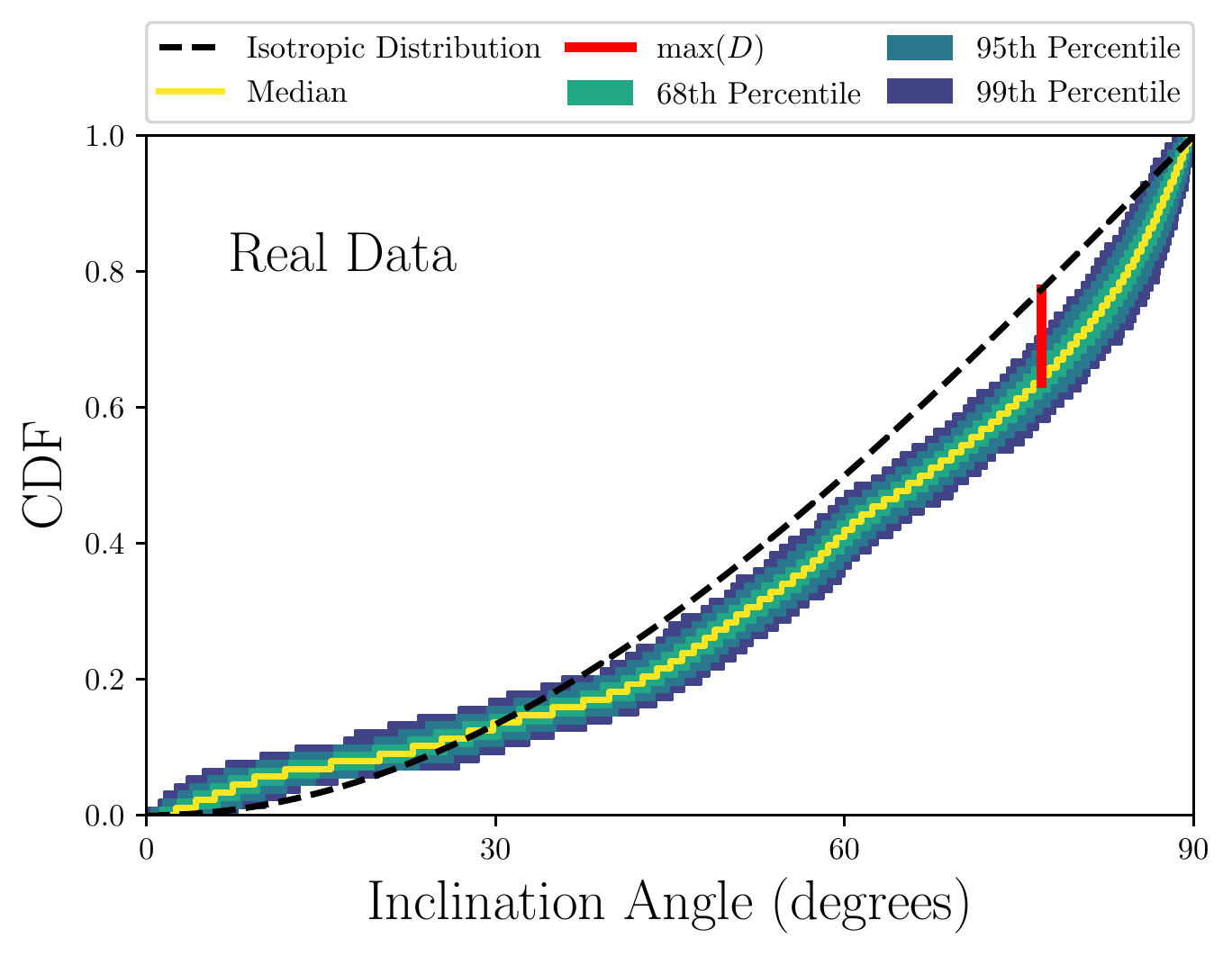}}
\caption{Empirical cumulative distribution functions (ECDFs) for the artificial data (top-panel) and real data (bottom panel) shown for 1000 draws (as described in text). The percentiles of the distribution are given along with the isotropic distribution shown by the black dashed line. The position of the largest deviation in each case is given by the red line.}
\label{fig:cdfs}
\end{figure}

First of all, we calculate the empirical cumulative distribution functions (ECDFs) of the artificial and real data by drawing an inclination angle from each of the inclination angle distribution for the stars we obtained a measurement for, constructing the ECDFs and then repeating this 1000 times, as shown in Fig~\ref{fig:cdfs}. Inspection of the ECDFs reveals the angles at which excesses lie compared to an underlying isotropic distribution. The gradient of the ECDF provides information about these excesses and where they occur, for example, at angles close to 0\degrees the ECDF of both datasets rises faster than the isotropic distribution reference indicating an excess and the subsequent plateau shows a lack of stars with respect to the isotropic distribution. For the intermediate angles there is good consistency between the distributions, indicating that the isotropic distribution represents the data well in this region. For high angles the ECDF of the datasets rise much faster than the isotropic distribution, again indicating an excess at those high angles. Note that this point is approximately where the largest deviation is found between the ECDFs and the isotropic distribution.

\begin{figure}
\centering
\centering
\subfloat[]{\includegraphics[width=0.45\textwidth]{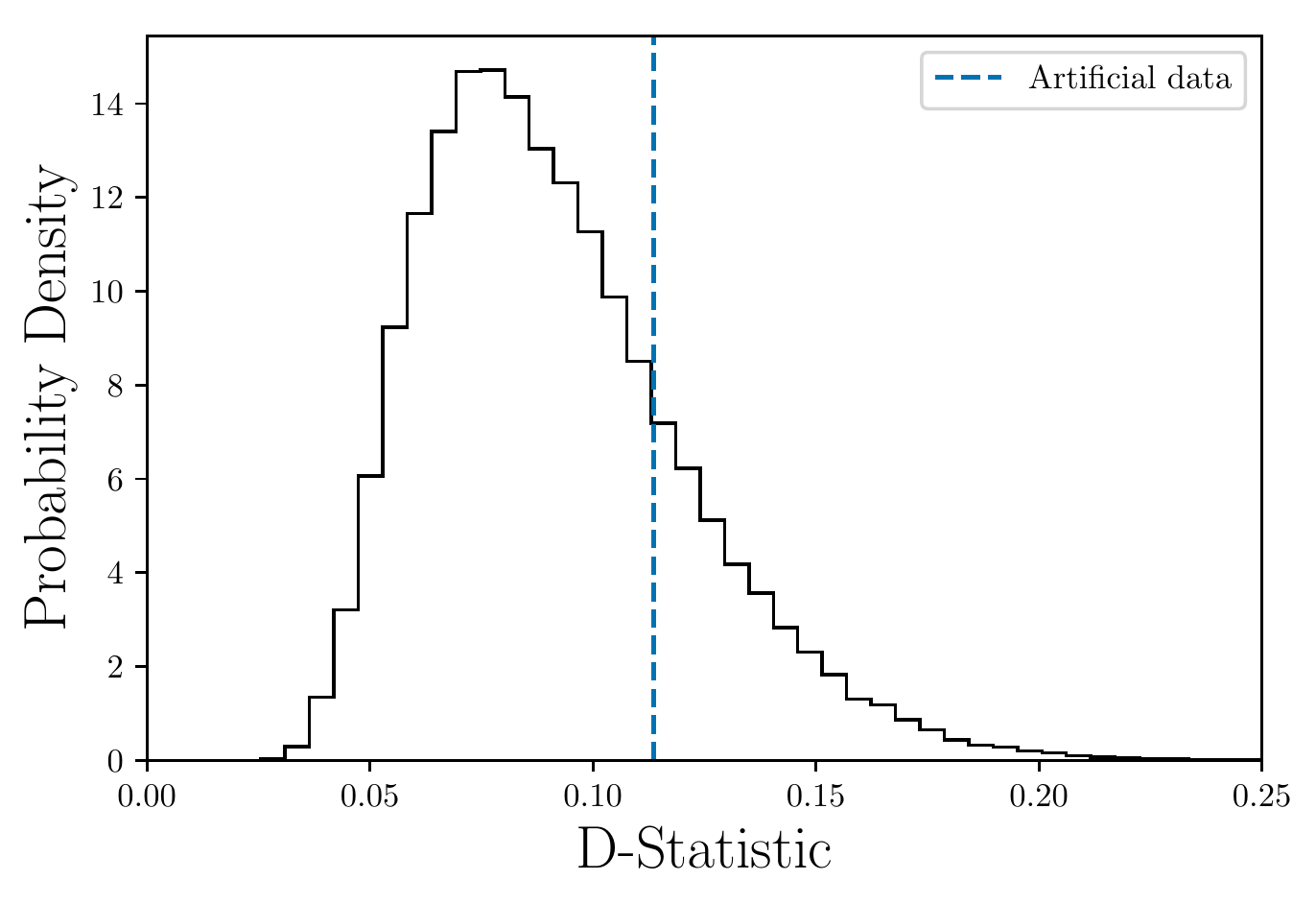}}
\vfill
\subfloat[]{\includegraphics[width=0.45\textwidth]{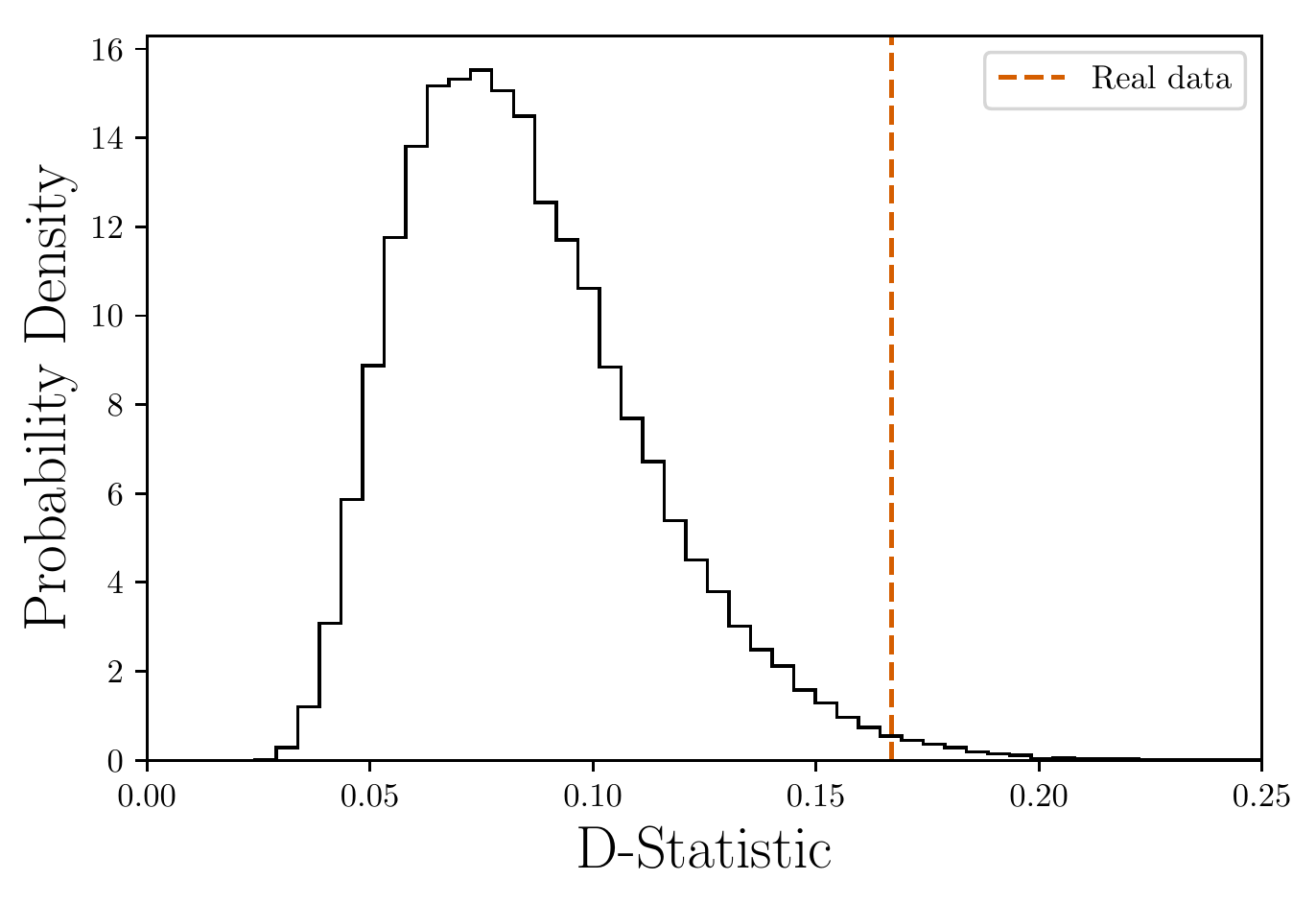}}
\caption{The distribution of the D-statistic for our sample size and $1\times10^{5}$ random draws. The top panel shows the distribution for the artificial data and the bottom panel shows the distribution for the real data, each time the D-statistic computed for each dataset is given by the dashed line.}
\label{fig:d_stat}
\end{figure}

For the K-S test, the isotropic distribution is taken as our reference distribution and what we wish to understand is what the expected distribution of stars would look like given the number of stars in our sample. To build our reference, we randomly draw an inclination angle (of the 89 stars in the case of the real data and 77 stars in the case of the artificial data) from an isotropic distribution and then calculate the empirical cumulative distribution function (ECDF) of those draws. We then compute the maximum deviation $D$ between the ECDF and the CDF of the isotropic distribution $P(i)\propto 1-\cos i$. This is repeated $1\times10^{5}$ times to build up a reference distribution of the maximum deviation expected for an isotropic distribution of our sample size, as shown in Fig.~\ref{fig:d_stat}.

The distribution of maximum deviation gives us the ability to assess where both the real data and artificial data lie in terms of consistency with an underlying isotropic distribution for the given sample size, this is highlighted in Fig~\ref{fig:cdfs} by the red lines. In the case of the artificial data the initial inclination angles were drawn from an isotropic distribution and so, as expected, the value of $D$ is consistent with this, lying at the 79th percentile. For the real data we do not know the underlying distribution, and the distribution of $D$ lies at the 99th percentile.

The discrepancy seen in the maximum deviation is unexpected. Given the way we have selected the stars in our sample there is no anticipated reason for the population distribution not to be consistent with isotropy. Before we can draw conclusions about the underlying population distribution, however, we must also consider that there are observational and population level biases that could well explain this discrepancy, such as the artificial simulations lacking physics that is present in the real data. These biases need only be small, and in individual cases can be less than the uncertainty on the inclination angle, but they are compounded when looking at the population as a whole. This motivates the need for a selection function. 

The fundamental quantity required to estimate the effect of the selection function on the population inference is the average detectability of a member of the population $\langle P_{\mathrm{det}}(i)\rangle$ \citep{2019MNRAS.486.1086M}
\begin{equation}
	\langle P_{\mathrm{det}}(i)\rangle = \int P_{\mathrm{det}}(i)p(i|\lambda)\mathrm{d}i,
\end{equation}
where subscript $\mathrm{det}$ denotes stars that have a detected inclination angle and $\lambda$ are the parameters of the population distribution (e.g. a modified Fisher distribution as used in this work, whose parameters can be inferred using an extension of the hierarchical inference to the population of stars). Due to the ``manual'' nature of the pipeline we use to estimate the inclination angle, estimating this quantity is beyond the scope of this paper; if the pipeline could be automated (e.g. \citealt{2018MNRAS.476.1470G}), then \cite{2019RNAAS...3e..66F} provides a simple way to estimate this quantity via a Monte-Carlo method with synthetic populations of stars with varying inclination angle and noise properties.

\section[]{Discussion}\label{sec:discussion}

\begin{figure}
 \includegraphics[width=0.45\textwidth]{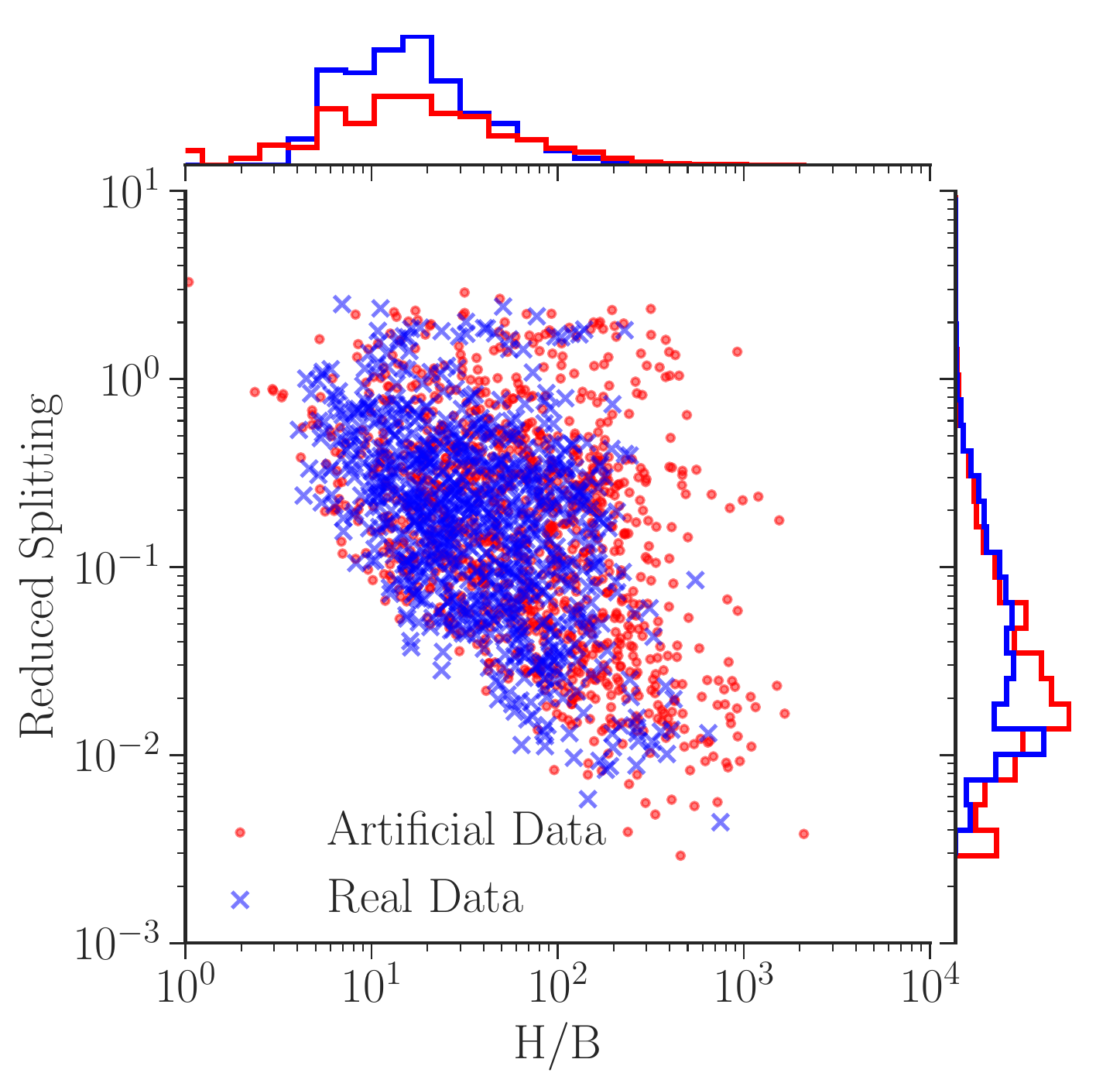}
\caption{A scatter plot showing the distributions of the real and artificial data (drawn from the isotropic input distribution) as a function of the inverse reduced splitting ($\Gamma/\nu_{\mathrm{s}}$) and signal-to-noise (given as the ratio of the mode height to the background). The marginalised densities for each parameter are shown in the histograms coloured according to the dataset.}
\label{fig: red_split_comp}
\end{figure}

In order for the hierarchical inference to work in the way presented here, we have made the assumption that all priors are separable and there are no dependencies of inclination angle on other parameters. This is weakly violated as the inclination angle is inferred through the relative mode heights. Therefore, if we were to make this method truly hierarchical, we would have to resample the amplitudes as well. This is, however, not necessary since both the artificial data and real data occupy the same region of parameter space in terms of signal-to-noise ratio and reduced splitting (the ratio of mode linewidth to the rotational splitting) as shown in Fig.~\ref{fig: red_split_comp}. Our ability to accurately retrieve the underlying inclination angle for the stars in our artificial sample allows us to place a constraint on the signal-to-noise ratio necessary to reliably extract the inclination angle. This can be placed at $\mathrm{H}/\mathrm{B}\gtrsim 4$, which can be seen in Fig.~\ref{fig: red_split_comp} as the cut where the density of points drops off greatly. 

\begin{figure}
 \includegraphics[width=0.45\textwidth]{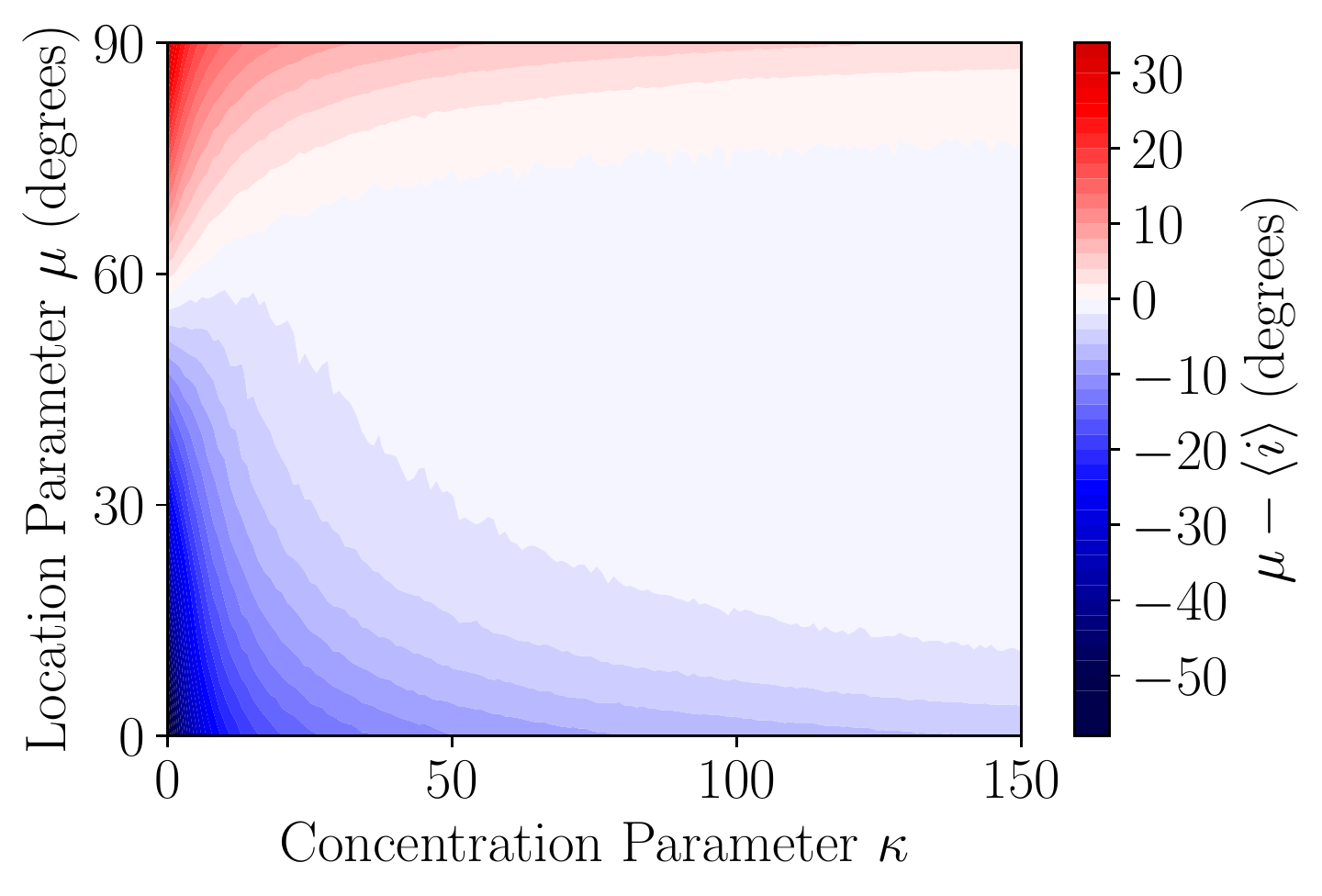}
\caption{A contour plot showing the difference between the location parameter in the updated Fisher distribution model and the mean of the distribution.}
\label{fig: model_diff}
\end{figure}

The model used for the analysis is a useful extension of the Fisher distribution, however there are a few issues that need to be addressed. The first is that there is no currently derivable analytical mean for the distribution. This can cause some issues with the interpretation of the location parameter $\mu$ of the model. In the case where $\kappa \rightarrow \infty$, $\mu$ will tend to the mean of the modified Fisher distribution, $\langle i\rangle$. For low $\kappa$ values this is not always the case due to $\mu$ not representing the mean of the distribution. We have assumed during the first part of the analysis that $\mu$ and $\langle i \rangle$ are representative of each other for large $\kappa$. This is best shown in Fig.~\ref{fig: model_diff} where the difference between $\mu$ and the mean of the distribution is shown for a range of $\kappa$ and $\mu$ values. The ranges of $\kappa$ spanned by each dataset are $20\lesssim \kappa \lesssim 145$, in the region where $\mu$ is representative of $\langle i\rangle$. This additional effect causes the hierarchical model to tend to underestimate stellar inclination angles towards zero and overestimate towards 90\degrees. Close to 0\degrees the location parameter is likely to underestimate the distribution mean whereas close to 90\degrees the opposite is the case (but to a much lesser degree). In an ideal case, the mean of the distribution would be the target parameter and Fig.~\ref{fig: model_diff} shows that for the majority of parameter space $\mu$ is indeed a valid alternative. 

In addition to the hierarchical inference there are also assumptions involved in the fitting of the modes which should be addressed. An important assumption that goes into the asteroseismic determination of the inclination angle is that there is equipartition of energy between modes of the same azimuthal order. If this assumption were not valid then for this random sample we would expect to not see isotropy in the inferred distribution. We cannot say that the assumption has been validated, although we can certainly say that there is no evidence to suggest that the assumption is not valid.

An observational bias that needs to be considered in the determination of the inclination angle is a 90$^{\circ}$ attractor, the fact that visually it is much easier to identify a mode in a singlet (close to 0 degrees) or doublet (close to 90 degrees) configuration for a given signal-to-noise ratio than a mode in a triplet configuration. This is due to the mode visibilities and how the power is distributed amongst the $m$ components of the mode. For a given underlying height, both the singlet and doublet configuration will have larger relative heights than the triplet observed at intermediate angles. In addition, assuming an underlying isotropic distribution, observing a star with an angle close to 0 degrees is rare which results in us being much more likely to observe a star near 90 degrees. This is the main reason for selecting stars with a high \numax value where mixed modes are much simpler to identify and so this effect is minimised in our sample. This will play an important role when analysing stars with much lower \numax values when the mixed mode identification becomes more difficult. The location of this bias at high inclination angles is in the same location as the excess we see in both datasets close to 90\degrees in Fig~\ref{fig:cdfs}.

The recent work of \cite{2018arXiv180507044K} highlighted a few potential sources of bias and warned against fitting oscillation modes individually, as we have done here, since this can lead to biases in the inferred inclination angle. This issue can be seen in the Monte-Carlo estimates of our artificial datasets, whereby there are significant biases. However, we have shown that through the use of hierarchical inference it is possible to extract the inclination angle reliably even when the oscillation modes are fitted individually. Fitting all of the oscillations at the same time using a heavily parametrised model could also potentially lead to biases that a set of individual fits would not be susceptible to. An example of this is Kepler-408 where \cite{2019AJ....157..137K} resolved conflicting estimates of the inclination angle obtained from global fits to the oscillation modes with a careful treatment of the granulation background. The advantage of our method here is that our estimate of the inclination angle is unaffected by the treatment of the granulation signal due to the fact that we fit modes individually where the background is locally flat. There are alternatives such as \cite{2018MNRAS.476.1470G} who fit all the modes at once making use of a maximum likelihood estimation and a novel peak-detection scheme. Either way, the important point is that care must be taken when extracting the inclination angle of any star. 

\cite{2018arXiv180507044K} also suggested that the p-dominated mixed modes should show severe blending because the ratio of the rotational splitting to the linewidth is less than 0.5. Fig.~\ref{fig: red_split_comp} shows the height-to-background ratio (HBR) as a function of the reduced splitting (which here is given by $\Gamma/\nu_{\mathrm{s}}$ rather than the reciprocal defined in \citealt{2018arXiv180507044K}) which shows that all but one of our modes across all of the data has $\Gamma/\nu_{\mathrm{s}} > 2$. This mode occurred in the artificial data set rather than the real data, and in the case of the real stars all modes lie below this limit showing that the severe blending is not an issue for these high $\nu_{\mathrm{max}}$ red-giant-branch stars.

\cite{2018A&A...616A..24G} and \cite{2018arXiv180708301M} have developed updated formalisms for describing the rotational splitting in red giant stars. The formalism presented in \cite{2018arXiv180708301M} is based upon integrating the function $\zeta$ over the frequency range between the $m=0$ and $m=\pm1$ components. The advantage of this prescription is that it can explain asymmetric rotational splittings without any need for extensions of the underlying theory. As a result, \cite{2018arXiv180708301M} reported asymmetric rotational splittings in red giant stars that exhibited fast rotation. In the case of slow rotators (such as our sample) this asymmetry is not observed. No asymmetric splittings were observed in our sample, however any small deviation from symmetry could manifest in small biases in the derived inclination angle. The degree of asymmetry however depends upon the mixing function $\zeta$ and so will be most extreme for the most g-dominated mixed modes, which in the high \numax regime we are working in are rare. Therefore the possibility of observing asymmetric rotational splittings in our sample is highly unlikely. When applying this method to stars with lower \numax values this will have to be an important consideration in the peakbagging and initial extraction of the inclination angle posteriors for each mode, rather than in the hierarchical method.

The possibility of asymmetric splittings is one such bias that could be present in our dataset due to assumptions made in the fitting process, such as assuming that the mode amplitudes are related by the deterministic expression given in Eqn~\ref{eqn: inc_leg}. If there are asymmetric splittings present (even slight) then the linewidths and heights of the individual azimuthal components could be affected in a similar way. Eqn~\ref{eqn: inc_leg} would therefore be correct for the underlying heights, but would then be affected by the mixing function $\zeta$. It is these sources of possible systematics that we account for by assuming an underlying inclination angle distribution for each star rather than a point estimate.

It has been suggested that searching for differences in the inclination angles inferred from p- and g-dominated mixed modes could yield evidence of core-envelope misalignment. For example, in \cite{2013Sci...342..331H} this was attempted but no such evidence was found. Unfortunately, due to the high \numax values of the stars in our sample we do not have enough g-dominated mixed modes to make this comparison.

Before asteroseismology can be used to infer full population distributions the complex selection function for the determination of the inclination angle needs to be determined. Whilst we can robustly determine the inclination angle for individual stars, there will be biases present that only become apparent when combining data at a population level, such as the relative ability to derive a given inclination angle (since some are easy to detect than others). This selection function will be dependent not only on the inclination angle, but also on the signal-to-noise ratio of the star. With the addition of a selection function then the hierarchical method presented in Section~\ref{sec: method1} can be readily extended to infer the population distribution.

\section[]{Conclusions}\label{sec:conclusions}

In this work, we have shown how asteroseismic estimates of the inclination angle of individual oscillation modes can be combined using a hierarchical Bayesian method to infer an underlying inclination angle for a single star. This has been applied to an artificial dataset that showed excellent agreement with the input inclination angles, and then applied to a sample of 123 high-$\nu_{\mathrm{max}}$ \emph{Kepler} red giants for which we report the inclination angles. The application of this method is not limited just to red-giant stars and can be readily applied to main-sequence stars to infer their inclination angles.

The hierarchical method proposed in this work could be extended to inferring the population distribution if we can derive and incorporate a selection function. This would alleviate population level biases that are not present in the individual estimates.

\section*{Acknowledgments}
We would like to thank the referee for their helpful comments and suggestions. We would also like to thank Nathalie Theme\ss l and Andr\'{e}s Garc\'{i}a Saravia Ortiz de Montellano for their very useful comments and discussions. An earlier version of this work appeared in J.S.K's PhD thesis (\url{http://etheses.bham.ac.uk/id/eprint/7658}). The research leading to the presented results has received funding from the European Research Council under the European Community's Seventh Framework Programme (FP7/2007-2013) / ERC grant agreement no. 338251 (StellarAges). This research has made use of the Exoplanet Follow-up Observation Program website, which is operated by the California Institute of Technology, under contract with the National Aeronautics and Space Administration under the Exoplanet Exploration Program. The authors acknowledge the support of the UK Science and Technology Facilities Council (STFC). Funding for the Stellar Astrophysics Centre is provided by the Danish National Research Foundation (Grant DNRF106). T. L. Campante acknowledges support from the European Union's Horizon 2020 research and innovation programme under the Marie Sk\l{}odowska-Curie grant agreement No. 792848 and from grant CIAAUP-12/2018-BPD.

This work made use of the following software: \texttt{corner} \citep{corner}, \texttt{emcee} \citep{emcee}, \texttt{Matplotlib} \citep{Hunter:2007}, \texttt{Numpy} \citep{doi:10.1109/MCSE.2011.37}, \texttt{Pandas} \citep{mckinney-proc-scipy-2010}, \texttt{scikit-learn} \citep{scikit-learn}, \texttt{Scipy} \citep{scipy}.


\bibliographystyle{mnras}
\bibliography{manuscript} 
\appendix

\section{Why is the isotropic prior uninformative?}\label{sec: why_uninform}

In any Bayesian analysis the choice of prior distributions is important, and not properly taking their effects into account can lead to poor inferences. It is common to use so-called ``uniformative'' priors whereby the prior has little-to-no effect on the parameter other than possibly providing constraints on the range the parameter could take. A common example is a uniform prior, which is assumed to be uninformative in many cases; however, this can in fact result in up-weighting the extremities of the prior range. What we aim to show here is that for the inclination angle, an isotropic prior distribution ($p(i) \propto \sin i$) informs our inference in a way that conforms with our physical interpretation of the underlying problem, resulting in it being an effective choice of ``uninformative'' prior distribution.

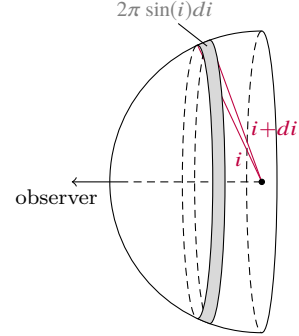
\begin{figure}
\center

\begin{tikzpicture}
  \coordinate (O) at (0,0);


  \draw[black] (0,2cm) arc [start angle = 90, end angle = 270,
    x radius = 2cm, y radius = 2cm];
  \draw[densely dashed] (0cm,2cm) arc [start angle = 90, end angle = 270,
    x radius = 2mm, y radius = 2cm];
  \draw (0cm,2cm) arc [start angle = 90, end angle = -90,
    x radius = 2mm, y radius = 2cm];

  \draw [purple] (O) to (110:2cm) node at (.17,.7) {\footnotesize $i$+$di$}; 
  \draw [purple] (O) to (115:2cm) node at (-.3,.3) {\footnotesize $i$}; 

  \draw[densely dashed] (O) -- (-1.8cm,0);
  \draw [->] (O)++(-1.8cm,0) -- (-2.5cm,0) node at (-2.7,-.2)  {\small observer};

  \filldraw (O) circle (1pt);

  \draw[densely dashed] (O) ++ (110:2cm) arc [start angle = 90, end angle = 270,
    x radius = 2mm, y radius = 2cm*sin(110)];
  \draw[densely dashed] (O) ++ (115:2cm) arc [start angle = 90, end angle = 270,
    x radius = 2mm, y radius = 2cm*sin(115)];

  \draw [fill = gray!30] (O) ++ (110:2cm) arc [start angle = 90, end angle = -90,
    x radius = 2mm, y radius = 2cm*sin(110)]
    arc [start angle = -110, end angle = -115,
    x radius = 2cm, y radius = 2cm]
    arc [start angle = -90, end angle = 90,
    x radius = 2mm, y radius = 2cm*sin(115)]
    arc [start angle = 115, end angle = 110,
    x radius = 2cm, y radius = 2cm];

  \draw (-1.1,2.1) -- (-0.72,1.8) node at (-1.27,2.27) [gray] {\footnotesize $2\pi\sin(i)di$};

\end{tikzpicture}
\caption{A demonstration of the geometry involved when interpreting the angle of inclination $i$.}
\label{fig: geom_angle}
\end{figure}

Inspecting the geometry of a system illuminates our prior knowledge of the inclination angle. Consider a star with a rotation axis that is inclined at an angle $i$ with respect to an observer, as shown in Fig.~\ref{fig: geom_angle}. With asteroseismology, we can only obtain information about $i$, defined for $0 \le i \le \pi/2$. By contrapositive, our observations are insensitive to the azimuthal angle $\varphi$ in the range $-\pi < \phi \leq \pi$. The likelihood of observing an inclination angle in the range $i$ to $i+di$ is proportional to the area of the corresponding annulus (Fig.~\ref{fig: geom_angle}). It is more likely for a star to be observed with an inclination angle near 90\degrees than near 0\degrees.

For an isotropic distribution, any direction on a unit hemisphere is equally probable. This constant probability density is given by $p=1/2\pi$ from normalizing over all solid angles of a hemisphere with area $2\pi$ steradians

We are insensitive to the azimuthal angle and only want to know the PDF of isotropic inclination angles, $p(i)$. The solid angle element can be decomposed in terms of $i$ and $\varphi$, giving $\mathrm{d}\Omega = \sin i \mathrm{d}i\mathrm{d}\varphi$. Therefore, the probability density $p$ can be expressed as a joint distribution over $i$ and $\varphi$
\begin{equation}
p(i, \varphi) = \frac{\sin i}{2\pi}.
\end{equation}
We want the distribution of $i$ rather than the joint distribution above, and so we marginalise over the azimuthal angle to get
\begin{equation}
p(i) = \int^{\pi}_{-\pi}p(i,\varphi)\mathrm{d}\varphi,
\end{equation}
resulting in the prior distribution for isotropic inclination angle defined for $i\in[0,\pi/2]$:
\begin{equation}
p(i) = \sin i.
\end{equation}

If we want to choose a prior that aligns with our current knowledge of the underlying process, i.e., that stars are randomly distributed in angle, then this isotropic prior is the most effective choice. A uniform prior, on the other hand, would give too much probability mass at low inclination angles, resulting in biases towards low inclination angles.

\section{Derivation of the marginalised likelihood}
\label{sec:like_deriv}

Given the hierarchical nature of the analysis, a natural starting point to deriving the marginalised likelihood in Eqn~\ref{eqn: new_like} is to write out the posterior distribution
\begin{equation}
p(\alpha,\theta|y)\propto p(\alpha, \theta)p(y|\alpha,\theta),
\label{eqn:hierarch_post}
\end{equation}
where $\alpha$ denotes the hyperparameters describing the form of the prior distribution and $\theta$ are the parameters of the fitted model. The joint prior distribution can be expanded to give
\begin{equation}
p(\alpha,\theta) = p(\alpha)p(\theta|\alpha),
\end{equation}
and so Eqn.~\ref{eqn:hierarch_post} becomes
\begin{equation}
p(\alpha,\theta|y)\propto p(\alpha)p(\theta|\alpha)p(y|\theta).
\end{equation}
An assumption has been made that the likelihood function $p(y|\theta,\alpha)$ has no explicit dependence on the hyperparameters $\alpha$. In other words, that the hyperparameters only affect the data through the parameters of the original model. When we say original model, or original fit, we mean the peakbagging model in which we extracted the inclination angle posterior distributions.

The parameters of the original fit, $\theta$, are nuisance parameters and we wish to marginalise them out to obtain our target posterior distribution $p(\alpha|y)$
\begin{equation}
p(\alpha|y) \propto \int p(\theta,\alpha|y)\mathrm{d}\theta,
\end{equation}
which thanks to the previous simplification leads to
\begin{equation}
p(\alpha|y) \propto p(\alpha)\int p(\theta|\alpha)p(y|\theta)\mathrm{d}\theta.
\end{equation}
The above integral is performed over all the parameters used in the original fitting and so will be high-dimensional meaning that a brute-force approach would be too computationally expensive. We can circumvent this by employing a trick (e.g. \citealt{MacKay:1999:IMC:308574.308666}) to obtain an importance resampled estimate
\begin{equation}
\int f(x)p(x)\mathrm{d}x = \int \frac{f(x)p(x)}{q(x)}q(x)\mathrm{d}x,
\end{equation}
which following the sampling approximation to integration can be given by
\begin{equation}
\int \frac{f(x)p(x)}{q(x)}q(x)\mathrm{d}x \approx \frac{1}{K}\sum_{k}\frac{f(x)^{(k)}p(x)^{(k)}}{q(x)^{(k)}}.
\label{eqn:importance}
\end{equation}
There are objections to the use of this approximation because of the seemingly arbitrary choice of the importance sampling distribution $q(x)$ \citep{10.2307/2348519}. However, in our case we can choose the posterior distributions themselves as the sampling distribution, i.e. $q(x) \propto p(\theta)p(y|\theta)$. Therefore Eqn.~\ref{eqn:importance} can be rewritten and simplified to give (using a subscript 0 to denote quantities from the original fits)
\begin{equation}
p(\alpha|y)\propto p(\alpha)\int\frac{p(\theta|\alpha)p_{0}(y|\theta)}{p_{0}(\theta)p_{0}(y|\theta)}\mathrm{d}\theta,
\end{equation}
which, when using the summation approximation, gives
\begin{equation}
p(\alpha|y)\propto p(\alpha)\frac{1}{K}\sum^{K}_{k=1}\frac{p(\theta|\alpha)^{(k)}}{p_{0}(\theta)^{(k)}}.
\label{eqn:marg_post}
\end{equation}

This applies for the case of a single star rather than the whole sample. To extend this to the full sample of $N$ stars we simply take the product over the marginalised likelihood in Eqn.~\ref{eqn:marg_post} giving, for the full sample,
\begin{equation}
p(\alpha|y)\propto p(\alpha)\prod^{N}_{n=1}\frac{1}{K}\sum^{K}_{k=1}\frac{p(\theta_{n}|\alpha)^{(k)}}{p_{0}(\theta_{n})^{(k)}},
\end{equation}
where we have assumed no covariances between the individual stars in our sample. This can also be simplified by using the assumption that there are no covariances between the parameters, such that $\theta$ can be replaced with the parameter of interest, the inclination angle $i$. This can also be applied to the quantity $p(\theta|\alpha)$, where we assume that $\alpha$ depends only on $i$ and so $p(\theta|\alpha)=f_{\alpha}(i)$, leading to the final form of the marginalised likelihood given in Eqn.~\ref{eqn: new_like}
\begin{equation}
p(y|\alpha)=\prod_{n=1}^{N}\frac{1}{K}\sum^{K}_{k=1}\frac{f_{\alpha}(i_{nk})}{p_{0}(i_{nk})}.
\end{equation}

\section{Derivation of normalising constant for modified Fisher distribution}
\label{sec: norm_deriv}

The use of numerical integration at every iteration when using MCMC is not ideal and so an analytical normalisation constant would help reduce the computation time and improve the efficiency of the hierarchical method.

The current unnormalised distribution is given by
\begin{equation}
f_{i}(i|\mu,\kappa) = \exp[\kappa\cos(i-\mu)]\sin i,
\end{equation}
and it is the integral of the exponential combined with the sine term that causes problems when trying to find an analytical solution.

To start we shall look at a very similar distribution to the modified Fisher distribution, the von-Mises distribution \citep{doi:10.1002/9780470627242.ch45}

\begin{equation}
f(x|\mu,\kappa)=\frac{\exp[\kappa\cos(x-\mu)]}{2\pi I_{0}(\kappa)},
\end{equation}
where $I_{0}(\kappa)$ is a modified Bessel function of the first kind of order zero.

There is a striking similarity between the von-Mises distribution and our modified Fisher distribution and enough of a resemblance to suggest that perhaps our distribution can also be normalised analytically. Another key hint is the fact that the cumulative distribution function (the integral of the probability density function) is also defined for the von-Mises distribution, suggesting that the integral of the numerator is possible.

Using section 9.6.34 of \cite{abramowitz+stegun}, the PDF of the von-Mises distribution can be approximated by making use of the identity

\begin{equation}
\exp(z\cos\theta) = I_{0}(z) + 2\sum^{\infty}_{k=1}I_{k}(z)\cos k\theta,
\end{equation}
which results in the PDF of the von-Mises distribution being rewritten as

\begin{equation}
f(x|\mu,\kappa)=\frac{1}{2\pi}\left\{1 + \frac{2}{I_{0}(\kappa)}\sum^{\infty}_{j=1}I_{j}(\kappa)\cos\left[j(x-\mu)\right]\right\}.
\end{equation}
The integral of the above equation is given by
\begin{equation}
\int f(x|\mu,\kappa) \mathrm{d}x = \frac{1}{2\pi}\left\{x + \frac{2}{I_{0}(\kappa)}\sum^{\infty}_{j=1}I_{j}(\kappa)\frac{\sin\left[j(x-\mu)\right]}{j}\right\}.
\end{equation}

The fact that the integral can be performed means that if we decompose the modified Fisher distribution in the manner above, then we should be able to integrate it in the same way. As a result, using the above decomposition the modified Fisher distribution becomes

\begin{equation}
f_{i}(i|\mu,\kappa) = \left\{I_{0}(\kappa) + 2\sum^{\infty}_{j=1}I_{j}(\kappa)\cos\left[j(i-\mu)\right]\right\}\sin i,
\end{equation}
which can be integrated to give
\begin{equation}
\int f_{i}(i|\mu,\kappa) \mathrm{d}x = \int\sin i\left\{I_{0}(\kappa) + 2\sum^{\infty}_{j=1}I_{j}(\kappa)\cos\left[j(i-\mu)\right]\right\}\mathrm{d}i.
\end{equation}
If we assume linearity within the integral then it can be split up to give
\begin{equation}
\int f_{i}(i|\mu,\kappa) \mathrm{d}i = \underbrace{I_{0}(\kappa)\int\sin i\mathrm{d}i}_\text{(1)} + \underbrace{2\int\sin i\sum^{\infty}_{j=1}I_{j}(\kappa)\cos\left[j(i-\mu)\right]\mathrm{d}i}_\text{(2)}.
\end{equation}
Now let us not forget that we are trying to find the normalisation constant and so

\begin{equation}
\int f_{i}(i|\mu,\kappa) \mathrm{d}i = \frac{1}{C},
\end{equation}
where $C$ is the normalisation constant.

Before we go any further with rearranging the equations to obtain $C$, let's first start by obtaining solutions to equations $(1)$ and $(2)$. We are integrating over the entirety of angle space, which in our case is $0$ to $\pi/2$ and so equation $(1)$ is very simple and the solution is just $I_{0}(\kappa)$. The solution to $(2)$ is a bit more involved. 

If we assume that the integral can go inside the summation along with the $\sin i$ term then we get

\begin{equation}
\begin{split}
2\sum^{\infty}_{j=1}I_{j}(\kappa)\int^{\pi/2}_{0}\sin i\cos\left[j(i-\mu)\right]\mathrm{d}i =\\
2\sum^{\infty}_{j=1}\frac{I_{j}(\kappa)}{j^{2}-1}\left[j\sin\left(\frac{1}{2}j(\pi-2\mu)\right) + \cos j\mu\right].
\end{split}
\end{equation}
The $j^{2}-1$ term is problematic as when $j=1$ the summation term is infinity and this should be avoided. We can instead expand out the summation and integrate each term individually, under the assumption that we only need the first $N$ terms of the infinite sum to approximate the underlying function.

\begin{equation}
\int^{\pi/2}_{0}\sin i\cos\left[j(i-\mu)\right]\mathrm{d}i = I_{1}(\kappa)\int^{\pi/2}_{0}\sin i\cos(i-\mu)\mathrm{d}i + ...,
\end{equation}
which we will now denote as $\varphi(\kappa, \mu)$. Therefore the normalising constant can be given as, 

\begin{equation}
C = \left\{I_{0}(\kappa) + 2\varphi(\kappa,\mu)\right\}^{-1},
\end{equation}
resulting in the normalised distribution

\begin{equation}
f_{i}(i|\mu,\kappa) = \left\{I_{0}(\kappa) + 2\varphi(\kappa,\mu)\right\}^{-1}\exp\left[\kappa\cos(i-\mu)\right]\sin i.
\end{equation}
Performing the summation in $\varphi(\kappa,\mu)$ up to $j\approx16$ is enough to achieve good precision with respect to numerical integration ($\sim 2$\% error for large $\kappa$, and ~1$\times10^{-14}$ for small $\kappa$).

\section{Derivation of cosine Fisher distributed angle with a location parameter}
\label{sec: model}

In order to derive the model distribution in $\cos i$ we follow the same line of analysis given in \cite{2014ApJ...796...47M}, i.e. we want to obtain the distribution of $f_{\cos i}$ given the distribution $f_{i}$. The following equation is used for $y = \cos i$

\begin{equation}
f_{Y}(y) = \left|\frac{\mathrm{d}}{\mathrm{d}y} g^{-1}(y)\right| f_{X}[g^{-1}(y)],
\label{eqn: prob_deriv}
\end{equation}
where $g^{-1}(y)$ is the inverse function of $y$ and we have dropped the summation over the number of solutions due to the fact that in the region of interest (0 to $\pi$) there is only one solution to $g^{-1}(y) = \arccos y$. The function $f_{X}$ is the probability distribution of the original data, which in our case is Eqn~\ref{eqn: model}. The first part of Eqn~\ref{eqn: prob_deriv} is given by

\begin{equation}
    \left|\frac{\mathrm{d}}{\mathrm{d}y} g^{-1}(y)\right| = \frac{1}{\sqrt{1 - y^2}},
\end{equation}
and the second part is
\begin{equation}
\begin{aligned}
    f_{X}\left[g^{-1}(y)\right] = {} & \left\{I_{0}(\kappa) + 2\varphi(\kappa,\mu)\right\}^{-1}\\
    & \exp\left[\kappa\cos\left(\arccos y - \mu\right)\right]\sin\left(\arccos y\right).
\end{aligned}
\end{equation}

The above equation can be simplified using trigonometric identities to give the following unnormalised distribution

\begin{equation}
f_{Y} (y | \mu,\kappa) = \left\{I_{0}(\kappa) + 2\varphi(\kappa,\mu)\right\}^{-1}\exp\left(\kappa y\cos\mu + \kappa\sqrt{1-y^{2}}\sin\mu\right).
\end{equation}

\section{Generating artificial red giant power spectra}
\label{sec: artif_spec}

In order to test the quality of the method and address any potential biases, an artificial dataset is created with known parameters. Both helio- and asteroseismic power spectra have been generated in previous works from the time-domain (e.g. \citealt{2006ESASP.624E..82C, 2008AN....329..549C}); however, in the case of red giants this is much more complicated due to the coupled nature of the modes \citep{2010Ap&SS.328..259D}. Therefore, instead of simulating red giant time series, we instead chose to generate the power spectrum directly by making extensive use of scaling relations. Currently, the artificial simulations are designed only to properly reproduce high $\nu_{\mathrm{max}}$ stars, but this can be extended in the future. The code used to generate these power spectra can be found at \url{https://github.com/jsk389/Artificial-red-giant-power-spectra}.

\subsection{Background spectrum}

The first part of the power spectrum to simulate is the granulation background and white noise components. The background profile adopted was that of model F from \cite{2014A&A...570A..41K}, which consists of two Harvey-like profiles describing granulation and a flat background due to the white noise. Whilst there is evidence to suggest the presence of a very low frequency granulation-like profile (most likely due to instrumental effects, \citealt{2014A&A...570A..41K}), this was not included in our simulations due to the lack of informed scaling relations to reproduce the required structure. 

Therefore, the background model is given as follows

\begin{equation}
    \mathcal{B}(\nu) = \eta^{2}\left(\sum^{2}_{i=1}\frac{\xi a_{i}^{2}/b_{i}}{1 + (\nu/b_{i})^{4}}\right) + W,
\end{equation}
where $\eta^{2}$ denotes the sinc-squared due to the sampling rate of the data (e.g. \citealt{2011ApJ...732...54C})
\begin{equation}
\eta^{2} = \mathrm{sinc}^{2}\left(\frac{\nu}{2\nu_{\mathrm{Nyq}}}\right),
\end{equation}
where $\nu_{\mathrm{Nyq}}$ is the Nyquist frequency of the observations. $a_{i}$ and $b_{i}$ are the respective amplitudes and characteristic frequencies of the $i$th granulation component, and $\xi = 2\sqrt{2}/\pi$ is a normalisation constant for the Harvey-like profiles with an exponent of 4.

\begin{table}
 \caption{The scaling relations describing the parameters used in the creation of the background spectrum. All are given of the form $\alpha\nu_{\mathrm{max}}^{\beta}$, where for the amplitude parameters $\alpha$ has units of $\mathrm{ppm}$ whereas for the characteristic frequency parameters it possesses units of $\mu$Hz.}
 \label{tab:backg}
 \begin{tabular}{lll}
  \hline
  Parameter & $\alpha$ & $\beta$\\
  \hline
  $a_{1,2}$ & 3382 & -0.609\\[2pt] 
  $b_{1}$ & 0.317 & 0.970\\[2pt]
  $b_{2}$ & 0.948 & 0.992\\[2pt]
  \hline
 \end{tabular}
\end{table}

The white noise was modelled according to the shot noise formulation given in \cite{2010ApJ...713L.120J} for \emph{Kepler} long-cadence data, and transforming it into the power spectrum domain following \cite{2011ApJ...732...54C}:

\begin{equation}
    W = 2\times 10^{6}\sigma^{2}\Delta t,
\end{equation}
where $\Delta t$ is the cadence of the instrument (29.4 minutes in the case of \emph{Kepler} long-cadence observations) and $\sigma$ is the root-mean-squared value of the noise \citep{2010ApJ...713L.120J}

\begin{equation}
    \sigma = 1\times 10^{6}\sqrt{c + 7\times10^{7}} / c,
\end{equation}
where $c$ is given by 

\begin{equation}
    c = 3.46\times10^{0.4(12 - K_{\mathrm{p}}) + 8},
\end{equation}
and $K_{\mathrm{p}}$ is the magnitude of the star in the \emph{Kepler} passband as given by the \emph{Kepler} Input Catalogue (KIC, \citealt{2011AJ....142..112B}).

\subsection{Mode frequencies}

Once the background of the power spectrum has been simulated, the focus can move to the oscillations themselves. The first property we are concerned with are their frequencies. This will be split into the cases of the radial and quadrupole modes, and then the mixed modes will be considered separately.

\subsubsection{Radial and quadrupole modes}

The asymptotic relation for mode frequencies is, to first order, given by \citep{1980ApJS...43..469T, 2011A&A...525L...9M, 2013A&A...550A.126M, 2015A&A...579A..84V}

\begin{equation}
    \nu_{n,\ell} = \left(n + \frac{\ell}{2} + \varepsilon + \frac{\alpha}{2}\left[n-n_{\mathrm{max}}\right]^{2}\right) \Delta\nu - \delta\nu_{n,\ell},
\label{eqn: norm_asymp}
\end{equation}
where $n$ is the radial order, $n_{\mathrm{max}}$ is the radial order at $\nu_{\mathrm{max}}$, $\ell$ is the degree, $\varepsilon$ is a phase term, $\Delta\nu$ is the large frequency separation, $\alpha$ describes the curvature in the $\Delta\nu$ and $\delta\nu_{n,\ell}$ is the small frequency separation. The above equation can produce frequencies for the radial ($\ell=0$) and quadrupole ($\ell=2$) modes for a given $\Delta\nu$ (given as an input to the simulation), $\varepsilon$, $\alpha$ and $\delta\nu_{n,\ell}$. In the case of radial modes the small separation is zero, and this simplifies the above equation. For modes of higher degree, the extra terms are described according to the following scaling relations \citep{2011A&A...525L...9M,2015A&A...579A..84V}

\begin{equation}
    \varepsilon = 0.634 + 0.546\log_{10}\Delta\nu,
\end{equation}

\begin{equation}
    \alpha = 0.015\Delta\nu^{-0.32},
\end{equation}
and finally \citep{2012ApJ...757..190C}

\begin{equation}
    \delta\nu_{n,\ell=2} = 0.121\Delta\nu + 0.035.
\end{equation}

As a result of the above scaling relations, we are also assuming that the observed $\ell=2$ modes are not mixed and pure pressure modes. \cite{2017A&A...605A..75D} demonstrated that observed $\ell=2$ modes can exhibit mixed behaviour, but again due the lack of expressions describing their properties for simplicity they are assumed to be pure pressure modes.

Due to the relatively small amount of power contributed by $\ell=3$ modes and the fact that there are no established scaling relations for their properties, these were also not included in the simulations. The scaling relations used in the simulations were also taken to be purely deterministic and any intrinsic scatter due to other parameters such as effective temperature or metallicity were not included.

\subsection{Dipole mixed modes}

The frequencies of mixed modes are a little more complicated to calculate given their mixed nature, and Eqn~\ref{eqn: norm_asymp} can no longer be used. Instead we follow the asymptotic expression given in \cite{2012A&A...540A.143M}

\begin{equation}
    \nu = \nu_{n_\mathrm{p},\ell=1} + \frac{\Delta\nu}{\pi}\arctan\left[q\tan\pi\left(\frac{1}{\Delta\Pi_{1}\nu}-\varepsilon_{\mathrm{g}}\right)\right],
\label{eqn: mixed_asymp}
\end{equation}
where $\nu_{n_\mathrm{p},\ell=1}$ is the nominal p-mode frequency, $q$ is the coupling factor, $\Delta\Pi_{1}$ is the $\ell=1$ period spacing and $\varepsilon_{\mathrm{g}}$ is a phase term. The mixed mode frequencies are obtained by finding the roots to Equation~\ref{eqn: mixed_asymp}. The nominal p-mode frequency is the frequency that the mode would take it is was purely acoustic, and so this can be approximated through the small frequency separation $\delta\nu_{01}$ \citep{2012ApJ...757..190C}:

\begin{equation}
    \delta\nu_{01} = \Delta\nu/2 + 0.109.
\end{equation}

For simplicity, both $q$ and $\varepsilon_{\mathrm{g}}$ are set to characteristic values for red giants, given by 0.2 and 0 respectively (as explained earlier in the text). Finally, the period spacing is approximated from $\Delta\nu$ by linearly interpolating the data given by \cite{2016A&A...588A..87V}.

\subsection{Mode Amplitudes}

Having calculated the mode frequencies, we can move onto the calculation of the mode amplitudes. The amplitudes of the $\ell=1$ and $\ell=2$ modes can be approximated from the radial mode amplitudes through the relative mode visibilities (e.g. \citealt{refId0, 2017ApJ...835..172L}). Let us start by calculating the amplitude of the radial mode at $\nu_{\mathrm{max}}$, $A_{\mathrm{max}}$, 

\begin{equation}
    A_{\mathrm{max}} = \sqrt{\frac{H_{\mathrm{env}}\Delta\nu}{\tilde{V}^{2}_{\mathrm{tot}}}},
\end{equation}
where $H_{\mathrm{env}}$ is the height of the Gaussian envelope that commonly describes the power excess in background fitting, and $\tilde{V}^{2}_{\mathrm{tot}}$ is the total visibility of the oscillations (taken as 3.16 for \emph{Kepler}; \citealt{2011A&A...531A.124B}). The height of the envelope is given by \citep{2012A&A...537A..30M}

\begin{equation}
    H_{\mathrm{env}} = 2.03\times10^{7}\nu_{\mathrm{max}}^{-2.38}.
\end{equation}

The radial mode amplitudes are then assumed to be distributed as the square root of a Gaussian with a given full-width at half maximum (FWHM), $\delta\nu_{\mathrm{env}}$,

\begin{equation}
    A_{\ell=0}(\nu) = A_{\mathrm{max}}\left[\exp\left(-\frac{(\nu-\nu_{\mathrm{max}})^{2}}{2\sigma^{2}}\right)\right]^{1/2},
\label{eqn: radial_amp}
\end{equation}
where $\delta\nu_{\mathrm{env}} = 2\sqrt{2\ln 2}\sigma$, which can be calculated following \citep{2012A&A...537A..30M}

\begin{equation}
    \delta\nu_{\mathrm{env}} = 0.66\nu_{\mathrm{max}}^{0.88}.
\end{equation}

The amplitudes of the $\ell=1$ and $\ell=2$ modes follow simply and can be calculated by evaluating Eqn~\ref{eqn: radial_amp} at the relevant frequencies (the nominal p-mode frequency for the mixed modes) and multiplying by the corresponding relative visibility. i.e.,

\begin{equation}
    A_{\ell=1,2} = \tilde{V}_{\ell=1,2}A_{\ell=0}(\nu_{\ell=1,2}).
\end{equation}

In order to take the effect of the changing mode linewidth into account for the mixed modes, we will use the height to calculate the mode profiles. The formulation is a variant of that used in \cite{2006MNRAS.371..935F} and can be seen in \cite{10.2307/j.ctt1vwmgmn}

\begin{equation}
    H = \frac{2\tilde{V}_{\ell=1,2}A_{\ell=1}^{2}(\nu_{\ell=1})}{\pi T\Gamma_{\ell=0}(\nu_{\ell=1}) + 2Q} ,
\end{equation}
where $\Gamma_{\ell=0}(\nu_{\ell=1})$ is the radial mode linewidth evaluated at the nominal p-mode frequency, $T$ is the length of the observing run and $Q$ is the ratio of the ineria of a non-radial mode ($I_{1}$) relative to the radial mode ($I_{0}$) evaluated at the same frequency. This can be approximated by \citep{10.2307/j.ctt1vwmgmn}

\begin{equation}
    Q \approx \frac{I_{1,\mathrm{g}} + I_{1,\mathrm{p}}}{I_{1,\mathrm{p}}},
\end{equation}
where $I_{1,\mathrm{p,g}}$ is the inertia in the respective p- or g-mode cavity. $Q$ is related to the mixing function $\zeta$ according to

\begin{equation}
    \zeta \approx \frac{Q-1}{Q},
\end{equation}
where $\zeta$ is the mixing function \citep{2015A&A...580A..96D}.

\subsection{Mode Linewidths}

It is known that the mode linewidth has a dependence on the effective temperature of the star (e.g. \citealt{2009A&A...500L..21C,2011A&A...529A..84B,2012ApJ...757..190C}) and considerable theoretical work has been performed to try and explain the complex mechanisms. Due to the complex dependencies, there is no simple scaling relation for the mode linewidth as a function of $\nu_{\mathrm{max}}$, therefore we borrow from our own data. All of the radial modes in our sample of red giants were peak-bagged using the Lorentzian formulation \citep{refId0} and their linewidths were extracted.

We slightly modify the model given in \cite{2014A&A...566A..20A} by assuming that the dip in the mode linewidths (given by $\nu_{\mathrm{dip}}$ in \citealt{2014A&A...566A..20A}) occurs at $\nu_{\mathrm{max}}$, giving

\begin{equation}
    \ln\Gamma = \alpha\ln\left(\nu/\nu_{\mathrm{max}}\right) + \ln\Gamma_{\alpha} - \left[\frac{\ln\Delta\Gamma_{\mathrm{dip}}}{1 + \left(\frac{2\ln\left(\nu/\nu_{\mathrm{max}}\right)}{\ln\left(W_{\mathrm{dip}}/\nu_{\mathrm{max}}\right)}\right)}\right],
\label{eqn: app_lws}
\end{equation}
where $\nu$ is the frequency of the mode, $\alpha$ is the exponent of the power law, $\Gamma_{\alpha}$ is the multiplicative factor in the power law, $\Delta\Gamma_{\mathrm{dip}}$ is the height (or depth) of the Lorentzian profile and $W_{\mathrm{dip}}$ is the width of the Lorentzian. This relation fits a power-law to the mode linewidth as a function of frequency in addition to a depression modelled by a Lorentzian. The mode linewidths were fitted as a function of reduced frequency (mode frequency divided by $\nu_{\mathrm{max}}$) using MCMC to sample the parameter space. The fitted parameters were then used to generate radial mode linewidths from Eqn~\ref{eqn: app_lws}.

\begin{figure}
 \includegraphics[width=0.45\textwidth]{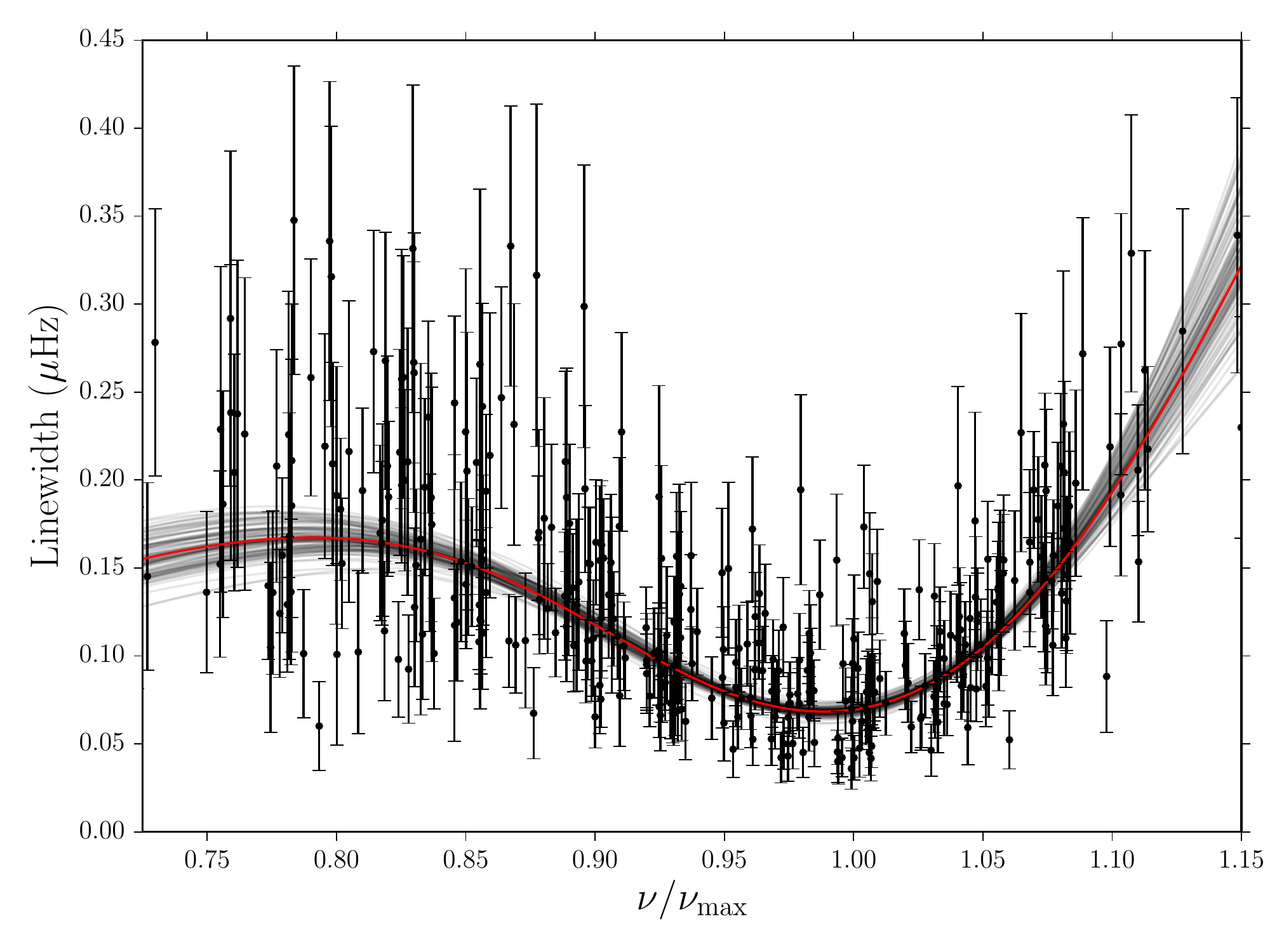}
\caption{Mode linewidths for the radial modes in our sample as a function of reduced frequency. The red solid line shows the fit to the data and overplotted are random draws from the posterior distributions of the parameters.}
\label{fig: lws}
\end{figure}

The $\ell=0$ and $\ell=2$ linewidths can be calculated by interpolating the above function at the respective frequencies\footnote{We have assumed that the properties of the $\ell=2$ modes are the same as the radial modes and that they are not mixed. Again, this is not strictly true since mixed $\ell=2$ modes have been observed \citep{2017A&A...605A..75D}, but due to there being little information about their properties we did not include mixed $\ell=2$ modes in the simulations.}. To calculate the mixed mode linewidths, we adopt the formulation given in \cite{2016AN....337..774D}

\begin{equation}
    \Gamma_{\ell=1} = \Gamma_{\ell=0}(\nu_{\mathrm{n}_{p},\ell=1})(1 - \zeta),
\end{equation}
where $\Gamma_{\ell=0}(\nu_{\mathrm{n}_{p},\ell=1})$ is the width of the radial mode evaluated at the nominal p-mode frequency.

\begin{table}
 \caption{Derived parameters from the fit to the radial mode linewidths.}
 \label{tab:lws}
 \begin{tabular}{lll}
  \hline
  Parameter & Value\\
  \hline
  $\alpha$ & $2.95^{+0.31}_{-0.32}$\\[2pt] 
  $\Gamma_{\alpha}$ ($\mu$Hz) & $0.52^{+0.11}_{-0.08}$\\[2pt]
  $\Delta\Gamma_{\mathrm{dip}}$ ($\mu$Hz) & $7.52^{+1.45}_{-1.16}$\\[2pt]
  $W_{\mathrm{dip}}$ & $0.79^{+0.02}_{-0.02}$\\[2pt]
  \hline
 \end{tabular}
\end{table}

Fig.~\ref{fig: red_split_comp} shows that the assumptions made during the simulations agree with the region of parameter space occupied by the low-luminosity red-giant branch stars used in this work.

\subsection{Rotational Splitting}

The final property to consider is the rotational splitting of the modes, for which we used the formulation given in \cite{2013A&A...549A..75G}

\begin{equation}
    \nu_{\mathrm{s}} = \left[\zeta(1-2\mathcal{R}) + 2\mathcal{R}\right]\nu_{\mathrm{s,max}},
\end{equation}
where $\mathcal{R}$ is the ratio of the average rotation rates of the envelope to the core, $\nu_{\mathrm{s,max}}$ is the maximum rotational splitting and $\zeta$ is again the mixing function given by \cite{2015A&A...580A..96D}. Due to the lack of any scaling-like relation for $\mathcal{R}$, we choose to adopt a value of $\mathcal{R}\approx0.01$, since this appears to reproduce the observed spectra to a good degree. 

Now that all of the mode properties have been defined, Eqn~\ref{model} can be used to generate the modes of oscillation.

\section{Inclination angles derived for real data}

\begin{table}
\caption{Derived inclination angles for the stars in the real sample. The median of the posterior distribution is given alongside the 68.3\% highest posterior density interval.}
\label{tab:real_incs}
\begin{tabularx}{\linewidth}{X X X X}
\toprule
 KIC & $i$ (degrees) & Positive uncertainty (degrees) & Negative uncertainty (degrees) \\
\midrule
  2158352 &                        83.95 &                            6.05 &                            2.91 \\
  2166709 &                        56.32 &                           11.48 &                           11.89 \\
  2308429 &                        29.75 &                            6.68 &                            6.35 \\
  2557441 &                         6.77 &                            3.17 &                            4.41 \\
  3111383 &                        19.30 &                            5.50 &                            4.82 \\
  3113213 &                        28.50 &                            7.78 &                            7.65 \\
  3223038 &                        84.73 &                            5.26 &                            2.25 \\
  3446775 &                        10.25 &                            4.28 &                            4.27 \\
  3531478 &                        46.76 &                            3.16 &                            3.08 \\
  3534077 &                        85.71 &                            4.29 &                            2.09 \\
  3634488 &                        70.69 &                            2.26 &                            2.35 \\
  3848387 &                        81.91 &                            7.66 &                            3.31 \\
  4042882 &                        56.81 &                            4.44 &                            4.32 \\
  4139632 &                        77.15 &                            3.93 &                            5.06 \\
  4141488 &                        66.46 &                            2.57 &                            2.55 \\
  4445966 &                        85.81 &                            4.19 &                            1.90 \\
  4445989 &                        55.64 &                            4.64 &                            4.55 \\
  4459359 &                        65.35 &                            3.04 &                            3.07 \\
  4482016 &                        73.47 &                            5.37 &                            6.05 \\
  4638467 &                        70.46 &                            5.75 &                            6.29 \\
  4646477 &                        20.87 &                            7.14 &                            5.96 \\
  4731138 &                        84.45 &                            3.31 &                            2.99 \\
  4738693 &                        80.89 &                            4.71 &                            5.03 \\
  4996676 &                         5.69 &                            2.55 &                            3.95 \\
  5025717 &                        80.56 &                            9.44 &                            3.60 \\
  5033397 &                        86.14 &                            3.86 &                            1.79 \\
  5115688 &                        44.57 &                            5.19 &                            5.06 \\
  5119742 &                        42.75 &                            3.00 &                            2.89 \\
  5198982 &                        44.92 &                            6.93 &                            6.59 \\
  5305291 &                        77.73 &                            3.64 &                            4.32 \\
  5428405 &                        86.04 &                            3.96 &                            1.95 \\
  5553307 &                        87.14 &                            2.86 &                            1.35 \\
  5623097 &                        59.50 &                            3.05 &                            2.83 \\
  5649129 &                        77.04 &                            7.60 &                            7.16 \\
  5731852 &                        87.71 &                            2.29 &                            1.22 \\
  5773365 &                        76.65 &                            4.33 &                            5.06 \\
  5879486 &                        81.41 &                            8.59 &                            3.78 \\
  5880144 &                        60.02 &                            6.42 &                            6.39 \\
  5961985 &                        83.79 &                            5.24 &                            2.96 \\
  6139471 &                        35.09 &                            6.49 &                            5.96 \\
  6208018 &                        87.64 &                            2.36 &                            1.15 \\
  6222530 &                        88.07 &                            1.93 &                            1.04 \\
  6307132 &                        72.00 &                            4.75 &                            5.51 \\
  6352407 &                        52.74 &                            2.84 &                            2.81 \\
\bottomrule
\end{tabularx}
\end{table}

\begin{table}
\contcaption{Continuation of Table~\ref{tab:real_incs}.}
\begin{tabularx}{0.9\linewidth}{X X X X}
\toprule
KIC & $i$ (degrees) & Positive uncertainty (degrees) & Negative uncertainty (degrees) \\
\midrule
  6776494 &                        60.22 &                            3.94 &                            3.75 \\
  6783217 &                        86.73 &                            3.27 &                            1.46 \\
  6924074 &                        57.62 &                            4.11 &                            4.05 \\
  6952783 &                        73.90 &                            3.52 &                            3.90 \\
  6964937 &                        76.94 &                            4.54 &                            5.62 \\
  7046554 &                        80.73 &                            5.96 &                            5.00 \\
  7468195 &                        37.18 &                            4.83 &                            4.48 \\
  7504619 &                        59.25 &                            2.39 &                            2.46 \\
  7584122 &                        49.12 &                            2.41 &                            2.38 \\
  7595722 &                        87.41 &                            2.59 &                            1.39 \\
  7693845 &                        84.89 &                            4.29 &                            2.52 \\
  7769544 &                        59.08 &                            2.66 &                            2.83 \\
  7898594 &                        59.82 &                            5.88 &                            5.64 \\
  8098454 &                        59.33 &                            2.97 &                            3.07 \\
  8107355 &                         6.24 &                            2.93 &                            4.59 \\
  8145017 &                        24.81 &                            5.65 &                            5.11 \\
  8192753 &                        49.41 &                            3.42 &                            3.15 \\
  8645227 &                         5.45 &                            2.52 &                            3.48 \\
  8827367 &                        53.94 &                            4.11 &                            3.84 \\
  8893299 &                        53.49 &                            3.97 &                            3.83 \\
  9145781 &                        88.57 &                            1.43 &                            0.82 \\
  9157260 &                        59.27 &                            2.73 &                            2.76 \\
  9219983 &                        27.43 &                            5.83 &                            4.75 \\
  9335457 &                        67.44 &                            2.54 &                            2.54 \\
  9418101 &                        83.18 &                            6.81 &                            2.66 \\
  9508218 &                        83.40 &                            6.60 &                            2.84 \\
  9814077 &                        85.72 &                            4.28 &                            2.04 \\
  9893437 &                        81.06 &                            4.26 &                            4.87 \\
  9896174 &                        82.06 &                            4.51 &                            4.64 \\
  9956184 &                        64.61 &                            3.47 &                            3.79 \\
 10198496 &                        68.72 &                            3.36 &                            3.34 \\
 10199289 &                        79.90 &                            2.65 &                            3.24 \\
 10353556 &                        29.53 &                            5.31 &                            4.66 \\
 10482211 &                        86.65 &                            3.35 &                            1.64 \\
 10581491 &                        74.23 &                            2.91 &                            3.43 \\
 10675916 &                        43.56 &                            3.76 &                            3.44 \\
 10734124 &                        42.45 &                            3.36 &                            3.28 \\
 10777735 &                        86.07 &                            3.93 &                            1.94 \\
 11015392 &                        87.48 &                            2.52 &                            1.26 \\
 11038809 &                        48.94 &                            3.55 &                            3.50 \\
 11043770 &                        73.44 &                            2.70 &                            3.17 \\
 11098411 &                         7.91 &                            3.75 &                            4.49 \\
 11852899 &                        50.66 &                            5.37 &                            5.02 \\
 12115374 &                        83.08 &                            5.83 &                            3.46 \\
 12203197 &                        46.75 &                            5.42 &                            5.06 \\
\bottomrule
\end{tabularx}
\end{table}
\end{document}